\documentclass[prc,superscriptaddress,showpacs,twocolumn,amssymb,amsmath,amsfonts,aps]{revtex4}
\usepackage[pdftex]{graphicx} 
\usepackage{epsfig}
\usepackage{dcolumn}  
\usepackage{longtable}
\usepackage{bm}       
\usepackage{subfigure}
\usepackage{multirow} %
\usepackage{epstopdf}
\newcommand\captionof[1]{\def\@captype{#1}\caption}
\newcommand\myreac{$\gamma p \rightarrow K^{+}\Sigma^0 $}
\newcommand\cmangle{$\cos \theta_{\mbox{\scriptsize{c.m.}}}^{K^+}$}
\newcommand\dsigma{$d\sigma / d \cos \theta_{\mbox{\scriptsize{c.m.}}}^{K^+}$}
\begin{document}
\newcommand*{\CMU}{Carnegie Mellon University, Pittsburgh, Pennsylvania 15213, USA}
\newcommand*{\CMUindex}{1}
\affiliation{\CMU}
\newcommand*{\WJ}{Washington \& Jefferson College, Washington, Pennsylvania 15301, USA}
\newcommand*{\WJindex}{2}
\affiliation{\WJ}
\newcommand*{\ANL}{Argonne National Laboratory, Argonne, Illinois 60441, USA}
\newcommand*{\ANLindex}{3}
\affiliation{\ANL}
\newcommand*{\ASU}{Arizona State University, Tempe, Arizona 85287, USA}
\newcommand*{\ASUindex}{4}
\affiliation{\ASU}
\newcommand*{\CSUDH}{California State University, Dominguez Hills, Carson, California 90747, USA}
\newcommand*{\CSUDHindex}{5}
\affiliation{\CSUDH}
\newcommand*{\CANISIUS}{Canisius College, Buffalo, New York 14208, USA}
\newcommand*{\CANISIUSindex}{6}
\affiliation{\CANISIUS}
\newcommand*{\CUA}{Catholic University of America, Washington, D.C. 20064, USA}
\newcommand*{\CUAindex}{7}
\affiliation{\CUA}
\newcommand*{\SACLAY}{CEA, Centre de Saclay, Irfu/Service de Physique Nucl\'eaire, 91191 Gif-sur-Yvette, France}
\newcommand*{\SACLAYindex}{8}
\affiliation{\SACLAY}
\newcommand*{\CNU}{Christopher Newport University, Newport News, Virginia 23606, USA}
\newcommand*{\CNUindex}{9}
\affiliation{\CNU}
\newcommand*{\UCONN}{University of Connecticut, Storrs, Connecticut 06269, USA}
\newcommand*{\UCONNindex}{10}
\affiliation{\UCONN}
\newcommand*{\EDINBURGH}{Edinburgh University, Edinburgh EH9 3JZ, United Kingdom}
\newcommand*{\EDINBURGHindex}{11}
\affiliation{\EDINBURGH}
\newcommand*{\FU}{Fairfield University, Fairfield Connecticut 06824, USA}
\newcommand*{\FUindex}{12}
\affiliation{\FU}
\newcommand*{\FIU}{Florida International University, Miami, Florida 33199, USA}
\newcommand*{\FIUindex}{13}
\affiliation{\FIU}
\newcommand*{\FSU}{Florida State University, Tallahassee, Florida 32306, USA}
\newcommand*{\FSUindex}{14}
\affiliation{\FSU}
\newcommand*{\GWU}{The George Washington University, Washington, DC 20052, USA}
\newcommand*{\GWUindex}{15}
\affiliation{\GWU}
\newcommand*{\ISU}{Idaho State University, Pocatello, Idaho 83209, USA}
\newcommand*{\ISUindex}{16}
\affiliation{\ISU}
\newcommand*{\INFNFR}{INFN, Laboratori Nazionali di Frascati, 00044 Frascati, Italy}
\newcommand*{\INFNFRindex}{17}
\affiliation{\INFNFR}
\newcommand*{\INFNGE}{INFN, Sezione di Genova, 16146 Genova, Italy}
\newcommand*{\INFNGEindex}{18}
\affiliation{\INFNGE}
\newcommand*{\INFNRO}{INFN, Sezione di Roma Tor Vergata, 00133 Rome, Italy}
\newcommand*{\INFNROindex}{19}
\affiliation{\INFNRO}
\newcommand*{\ORSAY}{Institut de Physique Nucl\'eaire ORSAY, Orsay, France}
\newcommand*{\ORSAYindex}{20}
\affiliation{\ORSAY}
\newcommand*{\ITEP}{Institute of Theoretical and Experimental Physics, Moscow, 117259, Russia}
\newcommand*{\ITEPindex}{21}
\affiliation{\ITEP}
\newcommand*{\JMU}{James Madison University, Harrisonburg, Virginia 22807, USA}
\newcommand*{\JMUindex}{22}
\affiliation{\JMU}
\newcommand*{\KNU}{Kyungpook National University, Daegu 702-701, Republic of Korea}
\newcommand*{\KNUindex}{23}
\affiliation{\KNU}
\newcommand*{\LPSC}{LPSC, Universite Joseph Fourier, CNRS/IN2P3, INPG, Grenoble, France
}
\newcommand*{\LPSCindex}{24}
\affiliation{\LPSC}
\newcommand*{\UNH}{University of New Hampshire, Durham, New Hampshire 03824, USA}
\newcommand*{\UNHindex}{25}
\affiliation{\UNH}
\newcommand*{\NSU}{Norfolk State University, Norfolk, Virginia 23504, USA}
\newcommand*{\NSUindex}{26}
\affiliation{\NSU}
\newcommand*{\OHIOU}{Ohio University, Athens, Ohio  45701, USA}
\newcommand*{\OHIOUindex}{27}
\affiliation{\OHIOU}
\newcommand*{\ODU}{Old Dominion University, Norfolk, Virginia 23529, USA}
\newcommand*{\ODUindex}{28}
\affiliation{\ODU}
\newcommand*{\RPI}{Rensselaer Polytechnic Institute, Troy, New York 12180, USA}
\newcommand*{\RPIindex}{29}
\affiliation{\RPI}
\newcommand*{\URICH}{University of Richmond, Richmond, Virginia 23173, USA}
\newcommand*{\URICHindex}{30}
\affiliation{\URICH}
\newcommand*{\ROMAII}{Universita' di Roma Tor Vergata, 00133 Rome Italy}
\newcommand*{\ROMAIIindex}{31}
\affiliation{\ROMAII}
\newcommand*{\MSU}{Skobeltsyn Nuclear Physics Institute, Skobeltsyn Nuclear Physics Institute, 119899 Moscow, Russia}
\newcommand*{\MSUindex}{32}
\affiliation{\MSU}
\newcommand*{\SCAROLINA}{University of South Carolina, Columbia, South Carolina 29208, USA}
\newcommand*{\SCAROLINAindex}{33}
\affiliation{\SCAROLINA}
\newcommand*{\JLAB}{Thomas Jefferson National Accelerator Facility, Newport News, Virginia 23606, USA}
\newcommand*{\JLABindex}{34}
\affiliation{\JLAB}
\newcommand*{\UNIONC}{Union College, Schenectady, New York 12308, USA}
\newcommand*{\UNIONCindex}{35}
\affiliation{\UNIONC}
\newcommand*{\UTFSM}{Universidad T\'{e}cnica Federico Santa Mar\'{i}a, Casilla 110-V Valpara\'{i}so, Chile}
\newcommand*{\UTFSMindex}{36}
\affiliation{\UTFSM}
\newcommand*{\GLASGOW}{University of Glasgow, Glasgow G12 8QQ, United Kingdom}
\newcommand*{\GLASGOWindex}{37}
\affiliation{\GLASGOW}
\newcommand*{\WM}{College of William and Mary, Williamsburg, Virginia 23187, USA}
\newcommand*{\WMindex}{38}
\affiliation{\WM}
\newcommand*{\YEREVAN}{Yerevan Physics Institute, 375036 Yerevan, Armenia}
\newcommand*{\YEREVANindex}{39}
\affiliation{\YEREVAN}
\newcommand*{\NOWSLAC}{Stanford University, Stanford, California 94305, USA}
\newcommand*{\NOWIMPERIAL}{Imperial College London, London SW7 2AZ, United Kingdom}
\newcommand*{\NOWGWU}{The George Washington University, Washington, DC 20052, USA}
\newcommand*{\NOWJLAB}{Thomas Jefferson National Accelerator Facility, Newport News, Virginia 23606, USA}
\newcommand*{\NOWLANL}{Los Alamos National Laborotory, New Mexico, USA}
\newcommand*{\NOWWM}{College of William and Mary, Williamsburg, Virginia 23187, USA}
%
%
\author {B.~Dey} 
\affiliation{\CMU}
\author {C.~A.~Meyer} 
\affiliation{\CMU}
\author {M.~Bellis} 
\altaffiliation[Current address: ]{\NOWSLAC}
\affiliation{\CMU}
\author{M.~E.~McCracken} 
\affiliation{\CMU}
\affiliation{\WJ}
\author{M.~Williams}
\altaffiliation[Current address: ]{\NOWIMPERIAL}
\affiliation{\CMU}
%
%
\author {K.~P.~Adhikari} 
\affiliation{\ODU}
\author {M.~Aghasyan} 
\affiliation{\INFNFR}
\author {M.~Anghinolfi} 
\affiliation{\INFNGE}
\author {J.~Ball} 
\affiliation{\SACLAY}
\author {M.~Battaglieri} 
\affiliation{\INFNGE}
% ...

\collaboration{The CLAS Collaboration}
\noaffiliation

%
%
%
%
%
%
%

%
\date{\today}
\title{Differential cross sections 
  and recoil polarizations for the reaction \myreac}
%
%
\begin{abstract} 
High-statistics measurements of differential cross sections and recoil polarizations for the reaction $\gamma p \rightarrow K^+ \Sigma^0$ have been obtained using the CLAS detector at Jefferson Lab. We cover center-of-mass energies ($\sqrt{s}$) from 1.69 to 2.84~GeV, with an extensive coverage in the $K^+$ production angle. Independent measurements were made using the $K^{+}p\pi^{-}$($\gamma$) and $K^{+}p$($\pi^-, \gamma$) final-state topologies, and were found to exhibit good agreement. Our differential cross sections show good agreement with earlier CLAS, SAPHIR and LEPS results, while offering better statistical precision and a 300-MeV increase in $\sqrt{s}$ coverage. Above $\sqrt{s} \approx 2.5$~GeV, $t$- and $u$-channel Regge scaling behavior can be seen at forward- and backward-angles, respectively. Our recoil polarization ($P_\Sigma$) measurements represent a substantial increase in kinematic coverage and enhanced precision over previous world data. At forward angles, we find that $P_\Sigma$ is of the same order of magnitude but opposite sign as $P_\Lambda$, in agreement with the static $SU(6)$ quark model prediction of $P_\Sigma \approx -P_\Lambda$. This expectation is violated in some mid- and backward-angle kinematic regimes, where $P_\Sigma$ and $P_\Lambda$ are of similar magnitudes but also have the same signs. In conjunction with several other meson photoproduction results recently published by CLAS, the present data will help constrain the partial-wave analyses being performed to search for missing baryon resonances. 
\end{abstract} 

\maketitle


\section{\label{section:intro}Introduction and Motivation}

Searches for missing baryon resonances in the strange sector have seen intense activity in recent years~\cite{mart_bennhold, janssen, rpr_lambda, rpr_sigma, sarantsev_2005, anisovich, bg_nikonov}, following theoretical predictions that several of these missing states have strong couplings to strange baryons~\cite{capstick-1998,Schumacher:1996ej}. For a recent review on baryons, see Ref.~\cite{Klempt:2009pi}. The $\gamma p \rightarrow K^+ \Sigma^0$ reaction promises to play an important role in this regard. It is closely related to the $\gamma p \rightarrow K^+\Lambda$ reaction but differs in an important aspect. Since $\Lambda$ is an isoscalar, only $I=\frac{1}{2}$ $N^{*}$ intermediate states can couple to $K^+\Lambda$ (isospin filter), while $\Sigma^0$ is an isovector, which allows it to couple to both $I=\frac{1}{2}$ $N^{*}$ and  $I=\frac{3}{2}$ $\Delta^{*}$ states. Coupled-channel analyses of these reactions are thus anticipated to be of special interest~\cite{usov-scholten, anisovich}.

The full scattering amplitude for $K^+ \Sigma^0$ photoproduction consists of eight complex production amplitudes corresponding to each of the two possible spin states of the incoming photon, target proton and outgoing hyperon. Parity relations reduce the number of independent complex amplitudes to four~\cite{fasano_tabakin_pwa_pol}, or the number of independent real observables to eight, one of them being the differential cross section (\dsigma) and the rest being a carefully chosen set of seven polarization observables~\cite{tabakin-1997}. Because of the self-analyzing nature of the hyperon ($\Lambda$ and $\Sigma^0$) decays, all seven of these polarization observables can be measured using different target and beam polarization configurations. It is thus possible to completely characterize the complex amplitude $A_{\gamma p \rightarrow K^+ \Sigma^0}$ from experimental observations. With an unpolarized beam and an unpolarized target however, only the differential cross sections and recoil polarizations ($P_\Sigma$) can be extracted.

Previous high statistics measurements for $\gamma p \rightarrow K^+ \Sigma^0$ have been made by the CLAS~\cite{mcnabb, bradford-dcs, bradford-cxcz}, SAPHIR~\cite{glander-saphir}, LEPS~\cite{kohri-leps, sumihama-leps}, and GRAAL~\cite{lleres-graal} Collaborations. The SAPHIR \dsigma measurements covered center-of-mass energies ($\sqrt{s}$) from threshold (1.69~GeV) to 2.4~GeV, while CLAS reached about 2.55~GeV. Agreement between the two data sets is fair, except in a few backward-angle bins where CLAS shows an enhancement around 2.2~GeV, whereas SAPHIR is more flat. The more recent LEPS experiment made precision measurements in the forward-angles for $\sqrt{s}$ from 1.93~GeV to 2.3~GeV, and is in overall fair to good agreement with CLAS and SAPHIR. World data on $K^+ \Sigma^0$ polarization is considerably more sparse, however. 

In this paper, we report new measurements of \dsigma and the recoil polarization for ${\gamma p \rightarrow K^+ \Sigma^0}$ using data taken at Jefferson Lab with the CLAS detector. We have utilized the decays $\Sigma^0 \rightarrow \Lambda \gamma$ and $\Lambda \rightarrow p \pi^-$, and have performed separate analyses for the final-state topologies $K^+p\pi^-$~($\gamma)$ and $K^+p$~$(\pi^-, \gamma)$, where the final-state particles that are not detected are shown in parentheses. Our \dsigma measurements cover center-of-mass (c.m.) energies from near production threshold ($1.69$~GeV) to $2.84$~GeV, in 10-MeV-wide bins. We also cover a wide angular range of $-0.95 \leq \cos \theta_{\mbox{\scriptsize{c.m.}}}^{K^+} \leq +0.95$, everywhere except the extreme forward- and backward-angles. Our angular binning is 0.1 in \cmangle. For the recoil polarization $P_{\Sigma}$, where available, we present results only from the $K^+p\pi^-$~($\gamma)$ topology, which allows us to preserve the spin-transfer information between the $\Sigma^0$ and the $\Lambda$ in the $\Sigma^0 \rightarrow \Lambda \gamma$ decay (see Sec.~\ref{section:pol}B). In the backward-angle and near-threshold bins, where statistics are extremely limited for the $K^+p\pi^-$~($\gamma)$ topology, our $P_{\Sigma}$ measurements are from the $K^+p$~($\pi^-, \gamma )$ topology instead.  For the most part, these results are in excellent agreement with the previous CLAS results~\cite{mcnabb}. In a few cases, the polarizations reported are not consistent with earlier measurements. These new measurements supersede the previous (lower statistics) CLAS results.

Theoretical and phenomenological model fits to previous world data incorporating intermediate resonances have typically suffered from the problem of ambiguity. That is, fits with different resonance contribution hypotheses gave comparable $\chi^2$ values. For example, it has long been known that there is a ``structure'' at around 1900~MeV in both the $K^+ \Lambda$ and $K^+ \Sigma^0$ differential cross sections. Mart and Bennhold~\cite{mart_bennhold} first pointed out that for $K^+ \Lambda$, this structure could be explained by the $D_{13}(1895)$ ``missing'' resonance. Janssen {\em et al.}~\cite{janssen} extended this model to $K^+ \Sigma^0$ and included the $S_{31}(1900)$ and $P_{31}(1910)$ resonances as well. They found that $D_{13}(1895)$ did not significantly improve the global fit quality for this channel and claimed that $u$-channel non-resonant processes could have a significant contribution instead. Various other groups~\cite{rpr_lambda, rpr_sigma, anisovich, mart_prc2006} have incorporated several other resonances, but it is fair to say that none of the models are conclusive.

Part of the problem lies in the fact that there has been very little data on the polarization observables, the aforementioned model fits being primarily based on cross section data. The problem is even more acute for the $\Sigma^0$, where the polarization is inherently ``diluted'' compared to the $\Lambda$ (this point is elaborated in Sec.~\ref{section:pol}B). High precision polarization measurements such as the current results are, therefore, much needed. With a finer binning in both energy and angular kinematic variables, several new features can be seen that were not apparent earlier. For instance, the SAPHIR paper~\cite{glander-saphir} noted that in accordance with quark model predictions, their data were consistent with $P_\Sigma \approx -P_\Lambda$, i.e., within their measurement uncertainties, the $\Lambda$ and $\Sigma^0$ recoil polarizations had comparable magnitudes and opposite signs. Our data show that this is not obeyed globally. The quark model assumes the same production mechanism for both hyperons, which can be badly broken in the resonance region if different resonances contribute to $\Lambda$ and $\Sigma^0$ production. Given that $\Delta^\ast$ states can couple only to $K^+ \Sigma^0$ and not $K^+ \Lambda$, this scenario cannot be ruled out.

Secondly, the treatment of non-resonant $t$- and $u$-channel contributions needs to be better understood. The quality of previous world data in the backward-angles has been found to be especially poor. The present data demonstrate a significant rise in the cross sections at backward-angles, pointing strongly to $u$-channel contributions, and therefore lend support to the same conclusion mentioned above from Ref.~\cite{janssen}.

These results have the largest kinematic coverage and represent the most precise measurements for this reaction that are available to date. In addition to the observables reported here, the LEPS Collaboration~\cite{kohri-leps} and the GRAAL Collaboration~\cite{lleres-graal} have also measured photon-beam asymmetries for this channel, while the FROST experiment at CLAS/JLab~\cite{frost} will measure several other single- and double-polarization observables. The present work is part of a larger program within the CLAS Collaboration to make precision measurements for several photoproduction 
channels~\cite{omega_prc, omega_pwa,eta_prc, klam_prc}, 
with the goal of performing a coupled-channel partial-wave analysis (PWA). 


\section{\label{section:exper}Experimental Setup}

The data used in this analysis were obtained using real photons produced via bremsstrahlung from a 4.023-GeV electron beam produced by the Continuous Electron Beam Accelerator Facility (CEBAF) at Jefferson Lab. The photons were energy tagged by measuring the momenta of the recoiling electrons with a dipole magnet and scintillator hodoscope system~\cite{sober}, resulting in a tagged photon energy range of $0.808$ to $3.811$~GeV for the current experiment. A separate set of scintillators was used to make accurate timing measurements. The photon energy resolution was about $0.1 \%$ of the incident beam energy and the timing resolution was 120~ps. These tagged photons were directed toward a $40$-cm-long cylindrical liquid-hydrogen cryotarget inside the CEBAF Large Angle Spectrometer (CLAS) detector system, which collected data events produced by scattering. Immediately surrounding the target cell was a ``start counter'' scintillator, used in the event trigger. 

Both the start counter and the CLAS detector were segmented into sectors with a six-fold azimuthal symmetry about the beam line. A non-uniform toroidal magnetic field with a peak strength of 1.8~T was used to bend the trajectories of charged particles and a series of drift chambers was used for charged particle tracking. In this manner, CLAS could detect charged particles and reconstruct their momenta over a large fraction of the $4 \pi$ solid angle. The overall momentum resolution of the detector was $\sim 0.5 \%$. A system of $\sim \!300$ scintillators placed outside the magnetic field and drift chamber regions provided timing information by measuring the time-of-flight (TOF) for each charged particle trajectory. A fast triggering and fast data-acquisition system (capable of running at $\sim 5$~kHz) allowed for operating at a photon flux of a few times $10^7$~photons/s. Further details of CLAS can be found in Ref.~\cite{mecking}.


\section{\label{section:data}Data}

The specific data set that we analyze in this work was collected in the summer of 2004, during the CLAS ``$g11a$'' run period. Roughly 20 billion triggers were recorded during this time, out of which only a small fraction corresponded to $K^+ \Sigma^0$ events. Each event trigger required a coincidence between the photon tagger Master OR (MOR) and the CLAS Level 1 trigger. For each charged particle to trigger individually, a coincidence between the TOF counter scintillator hit time and the start counter hit time for that particle was required. For the Level 1 trigger to fire, two particles in two different sectors of CLAS (``two-prong'' trigger) were required to trigger within a 150~ns coincidence time window. The final requirement was a coincidence between the tagger MOR and the start counter OR within a timing window of 15~ns. Also, only the first 40 tagger counters (corresponding to the higher end of the photon energy spectrum) were included in the trigger.

During offline processing, before any physics analysis began, the CLAS detector sub-systems had to be calibrated. This included determining the relative offsets between the photon tagger, start counter and TOF counter times, as well as calibration of the drift times in the drift chambers and the pulse heights of the TOF scintillators. Energy and momentum corrections were made for individual particles to account for their energy and momentum losses during passage through several layers of the detector sub-systems. Corrections were also made to the incident photon energy ($E_\gamma$) to account for mechanical sagging in the tagger hodoscope. A detailed discussion of the collection and calibration of this data set can be found in Refs.~\cite{will-thesis, my-thesis}.


\section{\label{sect:event_sel} Reaction topologies and Event Selection}

In the reaction $\gamma p \to K^+ \Sigma^0$, the $\Sigma^0$ subsequently decays as $\Sigma^0 \to \Lambda \gamma$ $100\%$ of the time. The $\Lambda$ further decays into $p \pi^-$ (charged mode) with a $64\%$ branching fraction and the rest of the time mostly into $n \pi^0$ (neutral mode). Since CLAS was optimized for detecting charged particles, we only use the charged decay mode of the $\Lambda$ in this analysis.
The ``three-track'' topology was then defined as $\gamma p \rightarrow K^+ p \pi^-$($\gamma$), where all three final state charged particles were detected and the outgoing photon was reconstructed from the missing momentum using a 1-constraint kinematic fit to zero total missing mass. This topology had a number of benefits. For example, the $\Lambda$ decay vertex could be reconstructed from proton and $\pi^-$ tracking information, leading to better energy loss corrections for these particles. Although the detection of the $\pi^-$ helped in the overall event reconstruction for the three-track topology, it also led to a reduction in the acceptance. Negatively charged particles (like $\pi^-$) were bent inwards toward the beam line, where the CLAS detector has its lowest acceptance.

To circumvent the above problem, we also examined the ``two-track'' topology, defined as $\gamma p \rightarrow K^+ p$($\pi^-, \gamma$), where only the $K^+$ and the proton were detected. With one less particle being detected, the acceptance was higher than in the three-track case, especially for the 
lower-energy and backward-angle regions. The two-track topology thus allowed for differential 
cross section measurements to be made from energies close to the $K^+ \Sigma^0$ production threshold and to cover almost the entire range in $\theta_{\mbox{\scriptsize{c.m.}}}^{K^+}$. Data samples for both topologies were then binned in 10-MeV-wide $\sqrt{s}$ bins for further analysis.

\begin{table}
  \centering
  \begin{tabular}{|l|c|c|} \hline \hline
    \multirow{2}{*}{Description} & \multicolumn{2}{|c|}{ Topology} \\
    \cline{2-3}
    &  \;\;\; $ K^+ p \pi^- (\gamma)$  \;\;\; &  \;\;\;  $K^+ p (\pi^-, \gamma)  \;\;\; $ \\ \hline
    Confidence level cut & \checkmark & -- \\ \hline
    $K^+ \Lambda$ removal cut & \checkmark & -- \\ \hline
    Timing cuts & \checkmark & \checkmark \\ \hline
    Total $MM$ cut  & -- & \checkmark \\ \hline  
    Fiducial cuts & \checkmark & \checkmark \\ \hline  
    \hline
    \end{tabular}
  \caption[]{\label{table:pid_cuts} List of cuts applied to the two topologies in this analysis. The confidence level and $K^+ \Lambda$ removal cuts used kinematic fitting that required the $\pi^-$ to be detected and could be applied to the three-track topology only. The total $MM$ cut came from an additional constraint on $MM(K^+ p)$ for the two-track data set. The fiducial volume cuts were applied to both topologies.
} 
\end{table}

Each event trigger recorded by CLAS consisted of one or more tagged photons. To begin the event selection process, at least two positively charged tracks were required to have been detected, as possible proton and $K^+$ candidates. The three-track topology required an extra negatively charged particle track, which was assumed to be a $\pi^-$. To minimize bias, all possible photon-particle combinations allowed by the given topology were taken to be candidate signal events. Event candidates with incorrectly assigned photons or particle hypotheses were removed by subsequent cuts. In the following sub-sections, we describe each of these event selection cuts one by one, referring the interested reader to Ref.~\cite{my-thesis} for further details. Since the two topologies followed significantly different analysis chains, to avoid confusion, we list the various cuts as applicable to each of the two- and three-track topologies in Table~\ref{table:pid_cuts}.

\subsection{Confidence level cut}

\begin{figure}
\centering
\includegraphics[angle=90,width=0.5\textwidth]{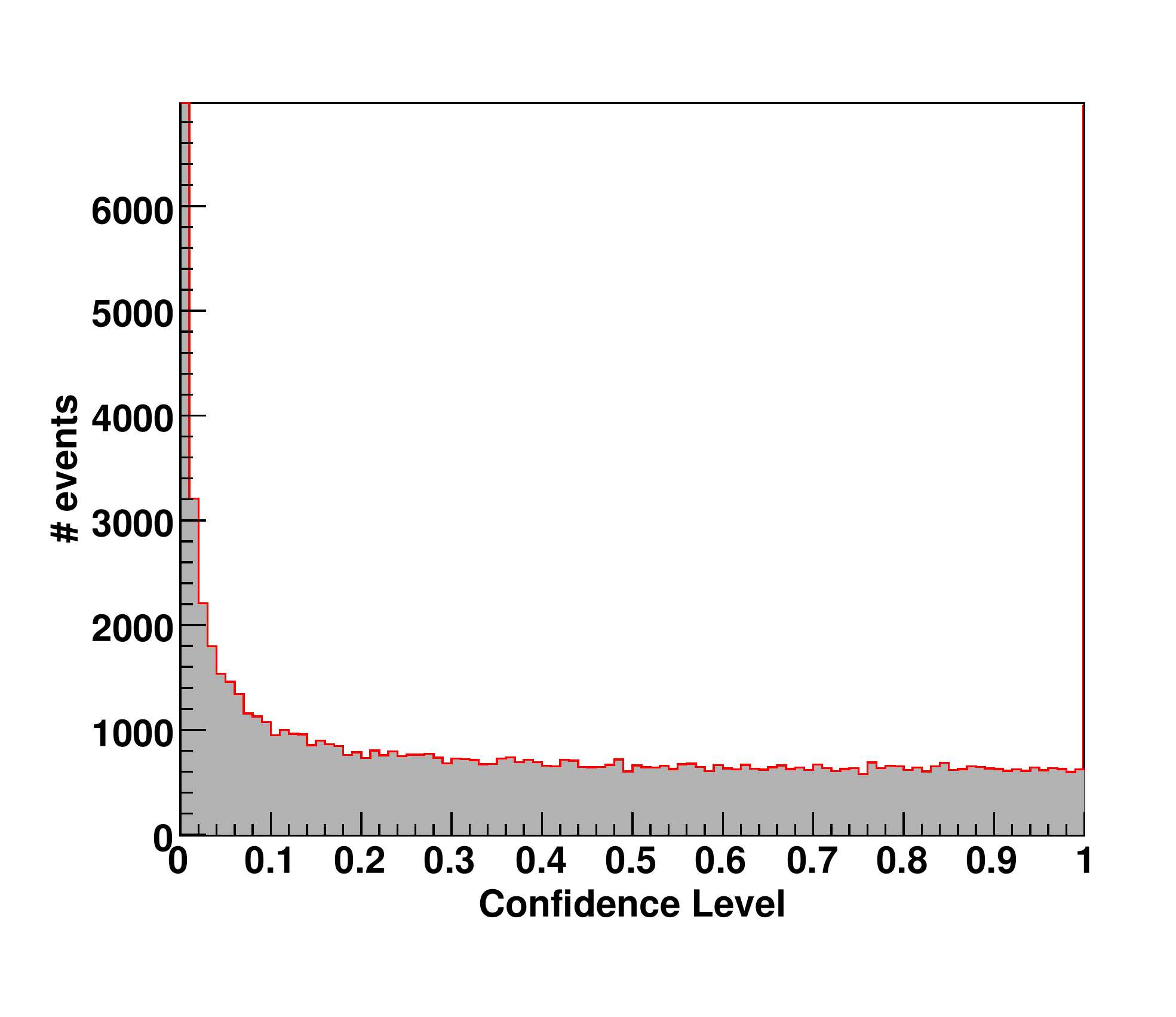}
\caption[]{(Color online) Confidence levels from a kinematic fit to the $\gamma p \to K^{+} p \pi^- (\gamma)$ reaction topology. The distribution was fairly flat above $\sim 10\%$, as expected for real signal events. Background events mostly occupied the region about zero. These were removed by placing a loose $1\%$ cut on the confidence level.}
\label{fig:confidence_level}
\end{figure}

Each event in the three-track data set was kinematically fit to an overall zero missing mass hypothesis for the undetected outgoing photon. This was a 1-constraint (1-$C$) kinematic fit. For every event recorded by CLAS, both the combinations ``$K^+:p:\pi^-$'' and ``$p:K^+:\pi^-$'' were treated as independent event hypotheses, where the three particle assignments corresponded to the three detected charged particles, two positively charged and one negatively charged. The kinematic fitter adjusted the momenta of each individual detected particle, while constraining the total missing mass to be zero. The shifts in the momenta, combined with the known detector resolution within the current experiment, gave a confidence level for the event to be the desired reaction. For a properly tuned kinematic fitter, background events have low confidence levels, while real signal events populate the confidence level with a flat distribution. The confidence level distribution is shown in Fig.~\ref{fig:confidence_level}. The peak near zero came from background events, which were removed by rejecting any event hypothesis with confidence level $< 1\%$. Above $\sim 10\%$, the distribution was fairly flat, as expected for real signal events.

\subsection{Timing cuts}

Track reconstruction through the different CLAS detector segments yielded both the momentum $\vec{p}$ and the path length $l$ from the reaction vertex to the TOF scintillator wall. The expected time-of-flight for a track hypothesized to be a particle of mass $m$ was then given by
\begin{equation}
  t_{\mbox{\scriptsize{exp}}} = \frac{l}{c}\sqrt{1 + \left(\frac{m}{p}\right)^2}.
\end{equation}
CLAS also measured the time-of-flight, $t_{\mbox{\scriptsize{meas}}}$, as the difference between the tagged photon's projected arrival time at the reaction vertex for the given event and the time the given particle track hits the TOF scintillator wall. The difference between these two time-of-flight calculations gave $\Delta tof = t_{\mbox{\scriptsize{meas}}} - t_{\mbox{\scriptsize{exp}}}$. For each track there was also a calculated mass $m_c$, given by
\begin{equation}
m_c =  \sqrt{\frac{p^2(1-\beta^2)}{\beta^2 c^2}},
\end{equation}
where $\beta = l/(ct_{\mbox{\scriptsize{meas}}})$. 

\begin{figure}
\begin{center}
\includegraphics[width=0.475\textwidth]{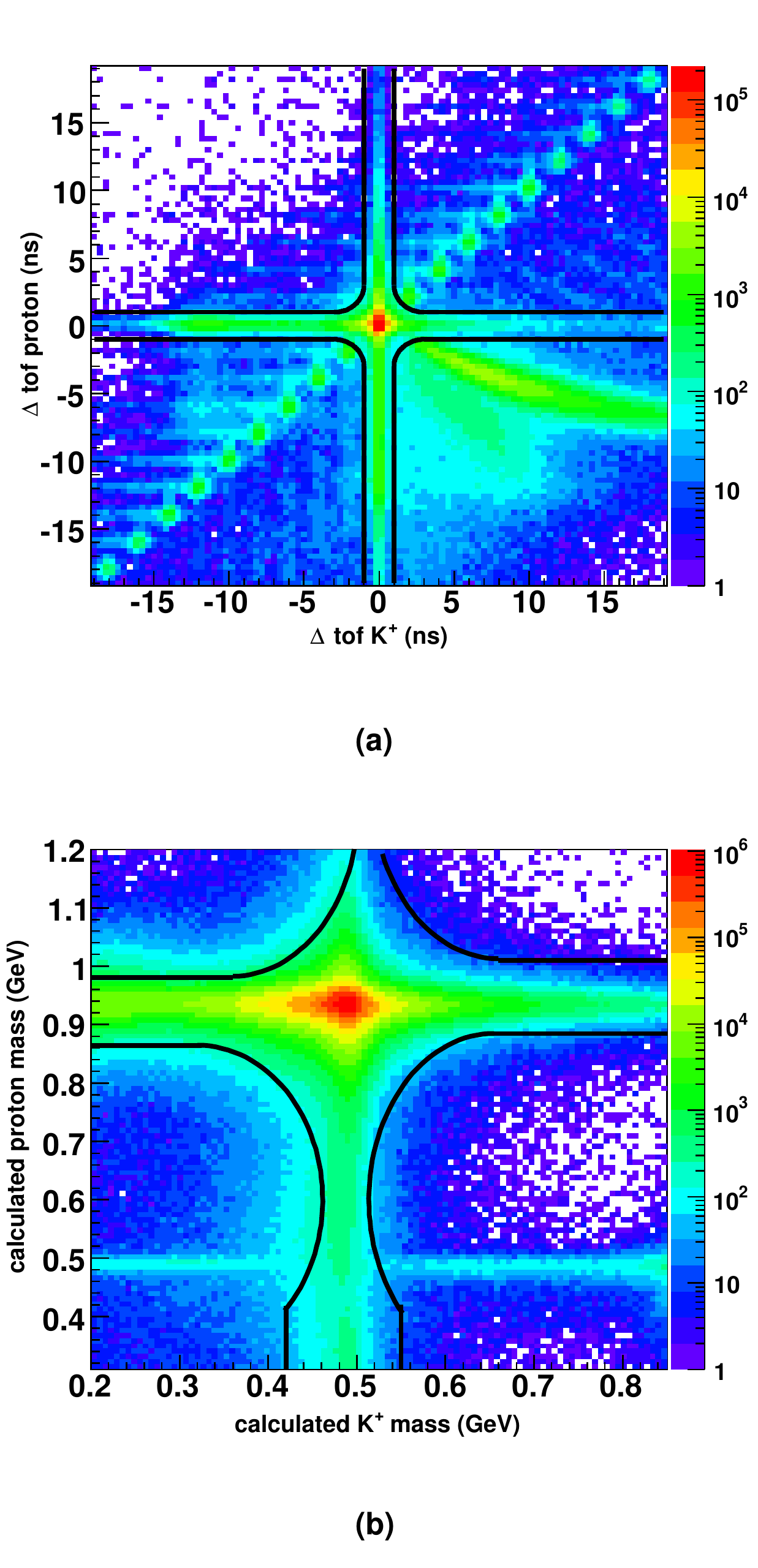}
\caption[]{(Color online) Timing cuts for background removal: (a) three-track topology, (b) two-track topology. Events lying outside the quadrant of black curves in both figures were removed from further analysis. Note the logarithmic scale for the intensity axes.}
\label{fig_timing_cuts}
\end{center}
\end{figure}

Timing information in the form of $\Delta tof$ or $m_c$ was used to place particle identification cuts on the proton and $K^+$ tracks for both the two- and three-track topologies. As mentioned earlier, for each pair of positively charged particle tracks, both ``$K^+:p$'' and ``$p:K^+$'' combinations were considered and treated as independent event hypotheses. The cuts are shown in Fig.~\ref{fig_timing_cuts}, where events outside the quadruplet of black curves were rejected. The clusters along the diagonal in Fig.~\ref{fig_timing_cuts}a, were due to accidental coincidences with events in different beam bursts corresponding to the 2~ns radio-frequency pulses used by the CEBAF electron accelerator. In general, our cuts were carefully tuned to keep signal loss at a minimum. The only possible exception was the upper-left cut boundary in Fig.~\ref{fig_timing_cuts}b that was kept tighter than the upper-right cut-boundary to reduce the very large pion background. 

\subsection{$K^+ \Lambda$ removal cut}

Since the $\Lambda$ and the $\Sigma^0$ are separated by only about 80~MeV in mass, some $K^+ \Lambda$ events invariably ``bled-in'' underneath the $K^+ \Sigma^0$ peak. This occurred most prominently in the kinematic regions where the lab angle between the $\Lambda$ and $\Sigma^0$ momenta (from $K^+ \Lambda$/$K^+ \Sigma^0$ production, respectively) was relatively small, typically for high energies and forward-angle scattering. 

For the three-track topology, it was possible to effectively remove the $K^+ \Lambda$ contamination using kinematic fitting. For this, every event hypothesis was kinematically fit to the topology $\gamma p \rightarrow K^+ p \pi^-$(nothing missing), which corresponded to the reaction $\gamma p \to K^+ \Lambda$. Since each component of the total 4-momentum was separately constrained to be zero, this was a highly constrained 4-$C$ fit. Events with a confidence level $>0.1\%$ from this 4-$C$ fit corresponded to $K^+\Lambda$ background and were removed from further analysis. 

\begin{figure}
\centering
\includegraphics[width=0.45\textwidth]{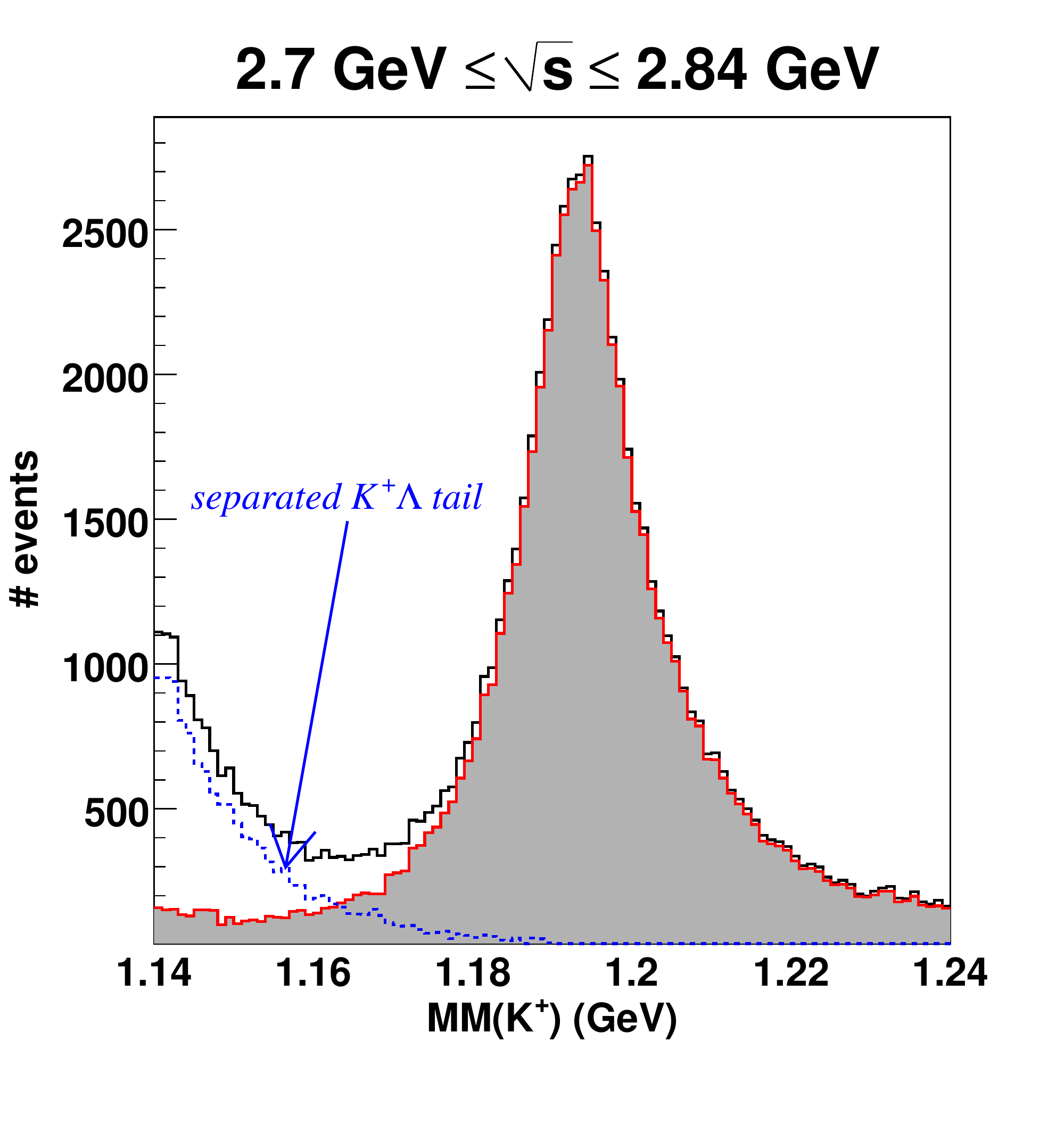}
\caption[]{(Color online) Effect of the $K^+ \Lambda$ removal cut: for the three-track topology, instead of fitting to $\gamma p \to K^+ p \pi^- (\gamma)$  for $K^+ \Sigma^0$ events, we can fit to $\gamma p \to K^+ p \pi^-$ (nothing missing) for $K^+ \Lambda$. By rejecting events with confidence level $ > 0.1 \%$ for the latter hypothesis, we can effectively remove the $K^+ \Lambda$ background tail (dotted histogram in blue). Only events in the shaded histogram were kept after this cut. 
}
\label{fig:lambda_removal}
\end{figure}

Fig.~\ref{fig:lambda_removal} shows the effect of this cut at high $\sqrt{s}$. The blue dotted histogram represents the unwanted $K^+\Lambda$ events that were leaking in previously and has two notable features. The first is the long tail from the $K^+\Lambda$ events with a peak at $MM(K^+) \approx 1.115$~GeV. The second is that it shows no sign of a peak around the $\Sigma^0$ mass, thereby implying that very few good signal events were removed by employing this cut. The shaded histogram represents the $K^+\Sigma^0$ events after $K^+ \Lambda$ background removal.   
 
Unfortunately, this cut required the $\pi^-$ to be detected, and thus could not be applied to the two-track data set. The remnant $K^+ \Lambda$ contamination for the two-track case was removed during signal-background separation (see Sec.~\ref{section:sig_bkgd}).   

\subsection{Total missing mass cut}

Consider the process $\Sigma^0 \to \Lambda \gamma \to p \pi^- \gamma$ from the perspective of a 3-body reaction. If the invariant mass $M(p \pi^-)$ is constrained to be $m_\Lambda$, then 3-body decay kinematics and the masses for the $\Sigma^0$, proton, $\pi^-$ and $\gamma$, lead to the bound
\begin{equation} 
0.176~\mbox{GeV} \leq M(\pi^- \gamma) \leq 0.251~\mbox{GeV}.
\label{eqn:dalitz}
\end{equation} 
In our reaction of interest, $M(\pi^- \gamma)$ also corresponds to $MM(K^+ p)$. For the three-track topology, since $\pi^-$ is explicitly detected and the outgoing photon is reconstructed via kinematic fitting, the above constraint is satisfied nominally. 

\begin{figure} 
  \centering                                                                                                    
  \includegraphics[width=0.45\textwidth]{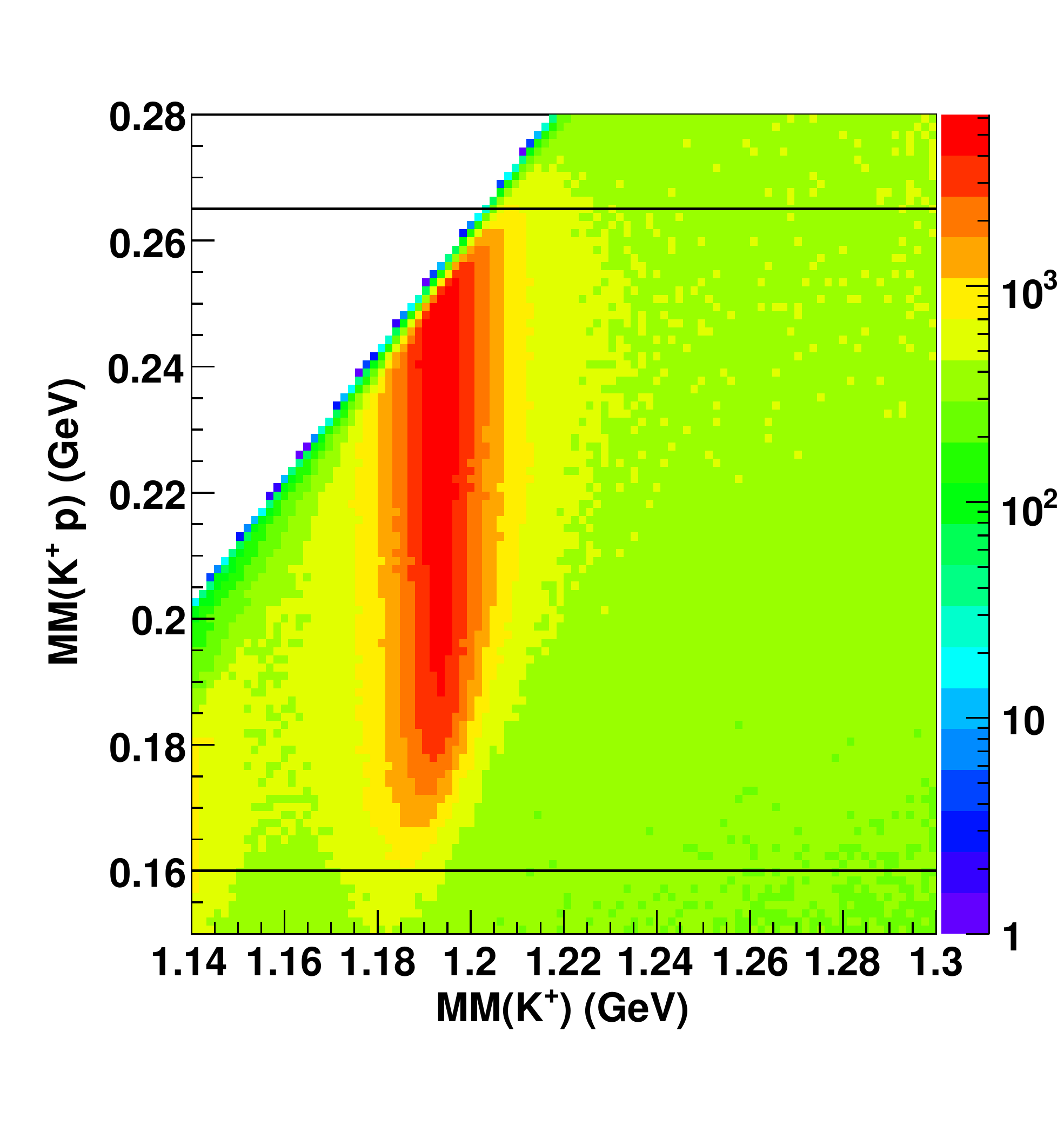}    
 \caption[]{(Color online) In the decay $\Sigma^0 \to \Lambda \gamma \to p \pi^- \gamma$, the invariant mass $M(\pi^- \gamma)$ is constrained to lie between 0.176 and 0.251~GeV. For the two-track topology, $M(\pi^- \gamma)$ corresponds to $MM(K^+ p)$, {\em i.e.}, the total missing mass. Therefore, events lying outside the region bounded by the two horizontal black lines were removed from further analysis.}                         
\label{fig:dalitz_cut}                                                                                      
\end{figure}

For the two-track topology, however, non-$K^+ \Sigma^0$ background events can lie outside the bounds given by Eq.~\ref{eqn:dalitz}. Since $MM(K^+ p)$ also corresponds to the total missing mass for the two-track data set, we place the following additional cut for this topology
\begin{equation} 
0.16~\mbox{GeV} \leq MM(K^+ p) \leq 0.256~\mbox{GeV}.
\label{eqn:dalitz1}
\end{equation} 
These upper and lower bounds were kept slightly looser than the values appearing in Eq.~\ref{eqn:dalitz} and are shown by the horizontal black lines in Fig.~\ref{fig:dalitz_cut}.

\subsection{Effectiveness of cuts}

The effectiveness of these cuts can be gauged by the percentage of ``signal'' events lost due to them. The $MM(K^+)$ distributions were fit to a Gaussian signal plus a quartic background function before and after placing the cuts. From this study, the loss in signal yields due to the cuts was estimated to be $\sim1.8\%$ for the two-track and $\sim0.62\%$ for the three-track topologies~\cite{my-thesis}. We quote these as the upper limits of the systematic uncertainties in our particle identification/event selection for this analysis.

\subsection{Fiducial cuts}

In addition to the above particle identification cuts, fiducial volume cuts were required to remove events belonging to regions where our understanding of the detector performance was relatively poor. These cuts were motivated by differences in an empirical efficiency calculation between the actual data and Monte Carlo which indicated discrepancies in the forward-angle region and at the boundaries of the six sectors of the CLAS detector due to edge effects in the drift chambers. Therefore, events with any particle trajectory falling near the sector boundary regions were removed. A $\phi_{\mbox{\scriptsize{lab}}}$-dependent cut on $\cos \theta_{\mbox{\scriptsize{lab}}}$ along with a hard cut at $\cos \theta_{\mbox{\scriptsize{lab}}} \geq 0.985$ removed extremely forward-going particles that coincided with the beam-dump direction. Localized inefficiencies within the CLAS detector volume such as inside the drift chambers were accounted for by placing trigger efficiency cuts on the Monte Carlo data as functions of $\phi_{\mbox{\scriptsize{lab}}}$, $\theta_{\mbox{\scriptsize{lab}}}$ and $|\vec{p}|$ for each particle track. Additional cuts were placed on backward-going tracks ($\cos \theta_{\mbox{\scriptsize{lab}}} \leq -0.5$). A minimum proton momentum cut at 375~MeV removed slow moving protons, whose energy losses were difficult to model in the detector simulation. Events with particles corresponding to poorly-performing TOF scintillator counters were removed as well.


\section{\label{section:sig_bkgd}Signal Background Separation}

\begin{figure}
  \centering
    \includegraphics[width=0.45\textwidth]{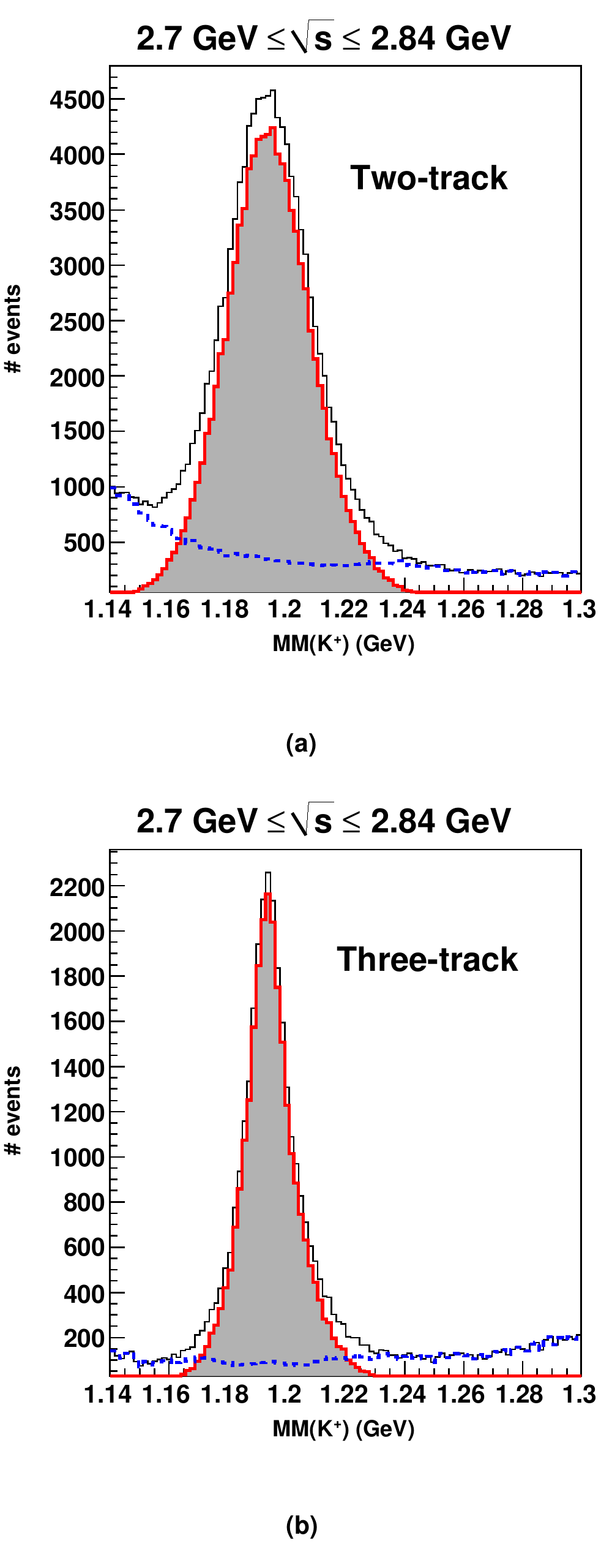}
\caption[]{\label{fig:sig_bkgd} (Color online) Signal background separation in the high $\sqrt{s}$ regions integrated over all angles for (a) two-track and (b) three-track topologies. The red (shaded) histograms are the data weighted by $Q$, representing the signal, while the blue (dotted) histograms are the data weighted by $(1-Q)$, representing the background. Note that the three-track data set has a smaller background from the $K^+ \Lambda$ ``tail'', because of the additional cut described in Sec.~\ref{sect:event_sel}B.
}
\end{figure}

The event selection cuts were very effective in cleaning the data sample for both topologies. Further removal of background, non-$K^+ \Sigma^0$ events, was affected by an event-based technique that sought to preserve correlations between all independent kinematic variables~\cite{my-thesis,jinst_williams}. The motivation behind this approach, as opposed to a more conventional sideband-subtraction method, was as follows.

For a reaction with multiple decays, such as in the present case, there are several independent kinematic variables (decay angles, for instance). To perform a background subtraction, one typically bins the data in a particular variable, such as the production angle $\theta_{\mbox{\scriptsize{c.m.}}}^{K^+}$. This is because the background level can vary widely within the range of the kinematic variable chosen. However, this binning in a single variable generally does not preserve correlations present in the other independent kinematic variables of interest. Therefore, one needs to bin the data in multiple kinematic variables, such that in any particular bin, the background level (both shape and size) remains roughly the same. Finally, the event-based fits using partial-wave amplitudes that we employ in the Monte Carlo to calculate the acceptance for the three-track topology (see Sec.~\ref{section:acc}) are specifically intended to reproduce the correlations present in the data. Therefore, we have adopted a more sophisticated approach for background separation.
 
To execute this technique for a given event, first, an $N_c$ number of ``closest neighbor'' events were chosen in the phase space of all independent kinematic variables. $N_c$ was typically of the order of a hundred. These $N_c +1$ events were then fitted to a Gaussian signal $s(m)$ plus a background function $b(m)$ using an event-based, unbinned, maximum likelihood method (the fit variable $m$ being $MM(K^+)$). Once the functions $s_i(m)$ and $b_i(m)$ had been obtained from this fit for the $i^{th}$ event, the event was assigned a signal quality factor $Q_i$ given by:
\begin{equation}
Q_i = s_i(m_i)/\left( s_i(m_i) + b_i(m_i)\right). 
\end{equation}
The $Q$-factor was then used to weigh the event's contribution for all subsequent calculations. In particular, the signal yield in a kinematic bin with $N$ events was obtained as
\begin{equation}
\mathcal{Y} = \sum\limits_{i}^{N} Q_i.
\end{equation}
Calculations were repeated using different forms of the background function and several different values of $N_c$, without any noticeable systematic shifts in the yields. The background function consisted of two parts. The first part was a Gaussian with a mean around the $\Lambda$ mass (the width was allowed to vary freely) -- this represented the $K^+ \Lambda$ ``bleed-in'' as was described in Sec.~\ref{sect:event_sel}C. The second part was modeled to represent the background from the high-mass end. Quadratic and Gaussian variants were tried out for this function. Trial values of $N_c$ were taken as 50, 100, 150, 200 and 300. We found that as long as $N_c$ was greater than $\approx 50$, the fits were stable. The final results presented here used $N_c = 200$. 

Fig.~\ref{fig:sig_bkgd} shows the results from applying this method for the two topologies for a given $\sqrt{s}$ bin. The background levels for the three-track topology varied from $< 5 \%$ at low energies to 5-10$\%$ at higher energies, but was generally found to be quite low. For the two-track topology, the background levels varied from 10-20$\%$, the backward-angles having a larger percentage of background than the forward-angles. The total data yields after all cuts and background separation were $\sim4.64$~million and $\sim0.65$~million for the two- and three-track topologies, respectively.


\section{\label{section:acc}Detector Acceptance}

Detector efficiency was modeled using GSIM, a GEANT-based simulation package of the CLAS detector. 300 million $\gamma p \rightarrow K^{+}\Sigma^0$ events were pseudo-randomly generated according to phase-space distributions and allowed to propagate through the simulation. An additional momentum smearing algorithm was applied to better match the resolution of the Monte Carlo with the real data. After processing, the ``raw'' ({\em i.e.}, original phase space generated) events yielded a set of ``accepted'' Monte Carlo events. The ``accepted'' Monte Carlo data then underwent the exact same series of event reconstruction, analysis cuts and energy-momentum correction steps as applied to the real data events. 

To account for the characteristics of the event trigger used in this experiment, two additional corrections went into the accepted Monte Carlo. The first of these corrections came from a trigger efficiency study using the $\gamma p \to p \pi^+ \pi^-$ channel. This study computed the probability that an individual particle trajectory did not fire the trigger, when the reaction kinematics strongly demanded (via total missing mass) that the particle should have been present. The average effect of this correction was found to be 5-6$\%$. 

The second of these corrections accounted for the macroscopic path length ($c \tau \sim 7.89$~cm) of the $\Lambda$. The start counter, which was included in the event trigger, surrounded the target cell at a distance of about 10 cm. In our reaction of interest, $\Lambda$ particles would decay into a proton and a $\pi^-$, and these daughter particles would fire the trigger (only charged particles could be detected by the start counter). Therefore, events where the $\Lambda$ decayed outside the start counter would not trigger the event readout. To correct for this, events in the accepted Monte Carlo data set, where the $\Lambda$ decay vertex lay outside the geometrical boundary of the start counter, were removed from further analysis. The probability of the daughter proton/$\pi^-$ tracks re-entering the start counter region was also studied and found to negligible. On average, the effect of this correction was about 5$\%$.  

To form a more accurate characterization of the detector acceptance pertaining to the kinematics of the reaction of interest, one should use a Monte Carlo event generator based on a physics model, instead of a simple phase-space generator. Typically, this is achieved in an iterative fashion; one starts with phase-space generated Monte Carlo events, extracts the differential cross sections, fits these cross sections to a model and uses the model to generate new Monte Carlo events for the next iteration. After several such iterations, the accepted Monte Carlo and data distributions are expected to resemble each other to a fair degree. 

However, the above procedure assumes that one has excellent control of the signal-background separation. For a complicated reaction with multiple decay angles, the detector acceptance can depend on several kinematic variables and it becomes more difficult to disentangle the effect of the detector acceptance on signal events from that on the background. Our signal-background separation procedure, as described in the previous section, specifically addresses this issue. By weighting every event by its $Q$-value, we are able to produce distributions of any particular kinematic variable that include only signal events. 

For the three-track topology, we expand the scattering amplitude $\mathcal{M}$ for the complete reaction chain $\gamma p \rightarrow K^+ \Sigma^0 \rightarrow K^+ p \pi^- \gamma$ in a basis of $s$-channel production amplitudes
\begin{equation}
\mathcal{M}_{\vec{m}}(\vec{x},\vec{\alpha}) \approx \linebreak 
\displaystyle \sum_{J=\frac{1}{2}}^{\frac{11}{2}} \sum_{P=\pm} A^{J^{P}}_{\vec{m}}(\vec{x},\vec{\alpha}),
\label{scat_exp}
\end{equation}
where $\vec{m}=\{m_{\gamma}, m_i, m_{\Sigma}, m_f, m_{\gamma f}\}$ denotes spin projections quantized along the beam direction for the incident photon, target proton, intermediate $\Sigma^0$, final-state proton and outgoing photon, respectively. The vector $\vec{x}$ represents the set of kinematic variables that completely describes the reaction, while the vector $\vec{\alpha}$ denotes a set of 34 fit parameters, estimated by a fit to the data distribution using the method of unbinned maximum likelihoods. The only assumption made here is that any distribution can be expanded in terms of partial waves (denoted by the spin-parity combination $J^P$). Ideally, one needs to use a ``complete'' basis of such $J^P$ waves, but we found that a ``large-enough'' ($J^P = \frac{1}{2}^{\pm}, \frac{3}{2}^{\pm}, \ldots ,  \frac{11}{2}^{\pm}$) set of waves was sufficient to fit the data very well. The $s$-channel $J^P$ waves were constructed using the relativistic Rarita-Schwinger formalism~\cite{rarita} and numerically evaluated using the {\tt qft++} software package~\cite{qft_pack}. A full description of the amplitude construction and fitting procedure can be found in Ref.~\cite{my-thesis}.

\begin{figure}
\includegraphics[width=0.45\textwidth]{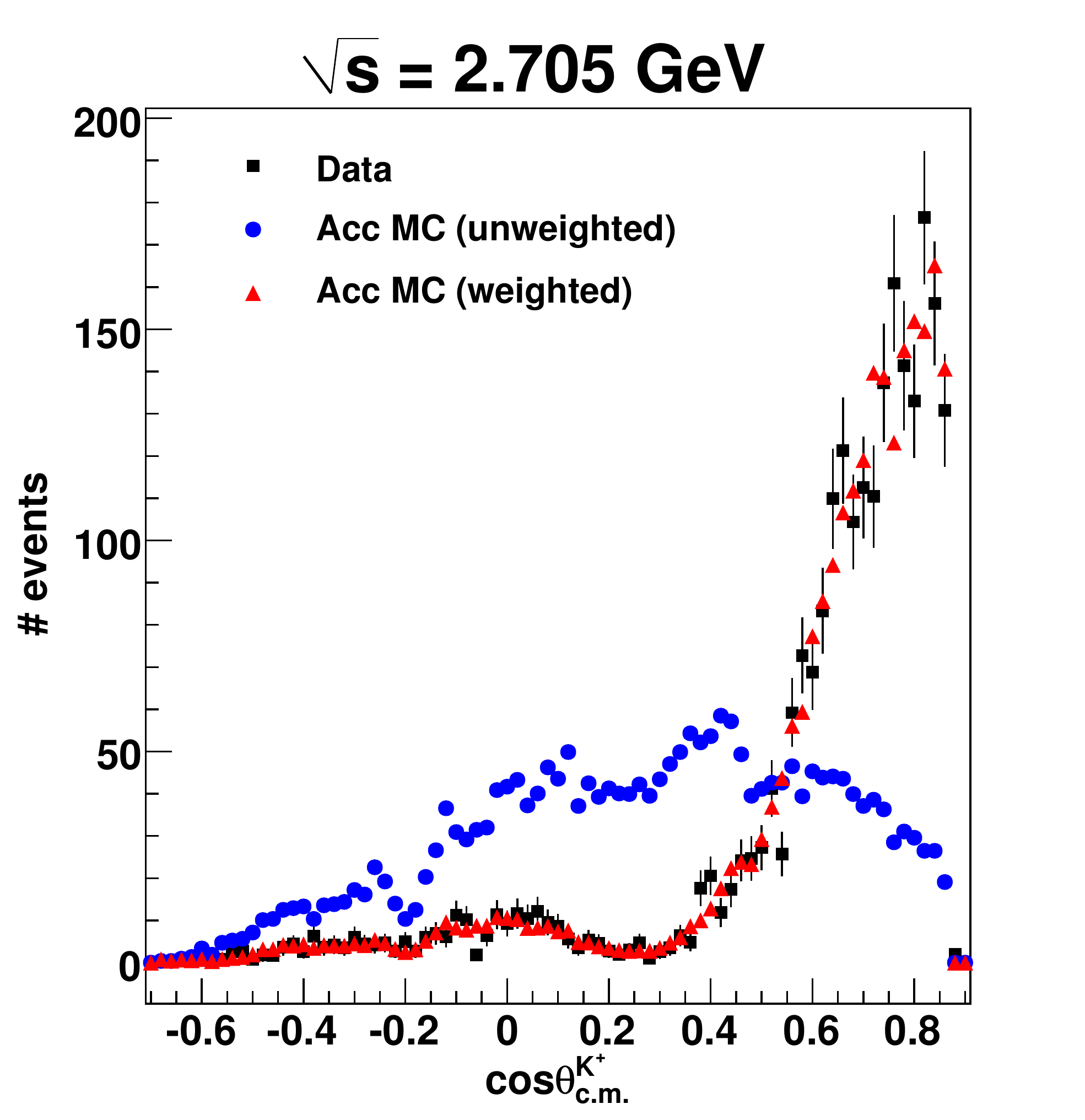}
\caption[]{\label{fig:weighted_acc}
  (Color online) 
  Shown are the \cmangle distributions for the data, accepted Monte Carlo and accepted Monte Carlo weighted by the PWA fit in the $\sqrt{s} = 2.705$~GeV bin for the three-track data set. Weighing by the fit results brings the weighted Monte Carlo distribution into excellent agreement with the real data. 
}
\end{figure}

Based on these fit results, each accepted Monte Carlo event was assigned a weight $I_{i}$ given by,
\begin{equation}
  I_{i} = \displaystyle \sum_{m_{\gamma},m_i,m_f,m_{\gamma f}}|\sum_{m_{\Sigma}} \mathcal{M}_{\vec{m}}(\vec{x}_{i},\vec{\alpha})|^{2},
\end{equation}
where we have coherently summed over the intermediate $\Sigma^0$ spins. The accepted Monte Carlo weighted by the fits matched the data in all physically significant distributions and correlations, as shown in Fig.~\ref{fig:weighted_acc} for the production angle. The detector acceptance as a function of the kinematic variables $\vec{x}$ was then calculated as
\begin{equation}
  \eta_{\,\mbox{\scriptsize{wtd}}}(\vec{x}) = \left( \displaystyle \sum_{i}^{N_{\mbox{\scriptsize{acc}}}} I_{i} \right) / \left( \displaystyle \sum_{j}^{N_{\mbox{\scriptsize{raw}}}}I_{j}\right),
\label{eqn_wtd_acceptance}
\end{equation}
where $N_{\mbox{\scriptsize{raw}}}$ and $N_{\mbox{\scriptsize{acc}}}$ denote the number of events in the given kinematic bin for the raw and the accepted Monte Carlo data sets, respectively.

The above procedure required knowledge of all final-state particle momenta. Since this was not available for the two-track topology, the acceptance in this case was calculated from the unweighted Monte Carlo intensities as
\begin{equation}
  \eta_{\,\mbox{\scriptsize{unwtd}}}(\vec{x}) = N_{\mbox{\scriptsize{acc}}}/N_{\mbox{\scriptsize{raw}}},
\label{eqn_unwtd_acceptance}
\end{equation}
where $N_{\mbox{\scriptsize{raw}}}$ and $N_{\mbox{\scriptsize{acc}}}$ are the same as in Eq.~\ref{eqn_wtd_acceptance}. This simpler expression was also used in a previous CLAS $K^+ \Sigma^0$ analysis~\cite{bradford-dcs}, where the effect of using a phase-space Monte Carlo generator, as opposed to a physics-model generator, was studied in detail. The conclusion from that study was that as long as the energy binning was fine enough such that the cross sections varied very little within each bin, the phase-space generator would give the correct acceptance. The previous CLAS analysis used $E_\gamma = 25$-MeV-wide bins, while our bins are even finer (10-MeV-wide in $\sqrt{s}$). Therefore, the above conclusion can be assumed to hold for the present case as well. In Sec.~\ref{section:results}A, we show that the differential cross sections using the two methods are in very good agreement.

\section{\label{section:norm}Normalization}

To calculate differential cross sections, the data yields were normalized by the photon flux and the target factors as
\begin{eqnarray}
  \frac{d\sigma}{d\cos \theta_{\mbox{\scriptsize{c.m.}}}^{K^+}}(\sqrt{s},\cos \theta_{\mbox{\scriptsize{c.m.}}}^{K^+}) = \left( \frac{A_{t}}{\mathcal{F}(\sqrt{s})\rho_{t}\ell_{t}N_{A}} \right) \times \nonumber \\
 \;\;\;\; \frac{\mathcal{Y}(\sqrt{s},\cos \theta_{\mbox{\scriptsize{c.m.}}}^{K^+})}{(\Delta \cos \theta_{\mbox{\scriptsize{c.m.}}}^{K^+})\eta(\sqrt{s},\cos \theta_{\mbox{\scriptsize{c.m.}}}^{K^+})},
\end{eqnarray}
where $A_{t}$, $\rho_{t}$, and $\ell_{t}$ were the target atomic weight, density and length, respectively, $N_{A}$ was the Avogadro constant, $\mathcal{F}(\sqrt{s})$ was the photon flux incident on the target for the given $\sqrt{s}$ bin, $\Delta\cos \theta_{\mbox{\scriptsize{c.m.}}}^{K^+}$ was the angular binning width, and $\mathcal{Y}(\sqrt{s},\cos \theta_{\mbox{\scriptsize{c.m.}}}^{K^+})$ and $\eta(\sqrt{s},\cos \theta_{\mbox{\scriptsize{c.m.}}}^{K^+})$ were the number of data events and the acceptance for the given kinematic bin, respectively.

Photon flux normalization for this analysis was carried out by measuring the rate of out-of-time electrons at the photon tagger, that is, hits that did not coincide with any event recorded by CLAS. Corrections were made to account for photon losses along the beam line and the detector dead-time.

A separate correction to the photon flux normalization was required to account for the fact that only the first two-thirds of the photon tagger counters (1-40) went into the trigger. ``Accidental'' events corresponding to tagger counters 41-61 could trigger if a simultaneous hit occurred in the lower (1-40) counters within the same time window. Such ``accidental'' events would be triggered as usual and recorded by CLAS just as any other ``normal'' event. However, the photon flux calculation would not incorporate the associated photon corresponding to an invalid tagger counter. By utilizing the trigger rates in counters 1-40 and assuming a Poisson distribution for the probability of occurrence of such ``accidental'' events, we were able to correct for this feature. The boundary between the $40^{th}$ and $41^{st}$ counters corresponded to the energy bin $\sqrt{s} =~1.955$~GeV, which had an unreliable flux due to this correction. As well, faulty tagger electronics prevented accurate electron rate measurements for photons in the energy bins $\sqrt{s}$~=~2.735~and~2.745~GeV~\cite{my-thesis}. Differential cross sections are therefore not reported at these three energies. However, polarization measurements do not depend on flux normalizations and are reported in these three bins.


\section{\label{section:syst}Uncertainties}

The statistical uncertainties for the differential cross sections were comprised of the uncertainty in the data yield and the acceptance calculation. For the $i^{th}$ event, the covariance matrix from the signal-background fit described in Sec.\ref{section:sig_bkgd} gave the uncertainty $\sigma_{Q_i}$ in our estimate of the signal quality factor $Q_i$. Summing up these uncertainties, assuming 100$\%$ correlation for events in a given $(\sqrt{s}, \cos \theta_{\mbox{\scriptsize{c.m.}}}^{K^+})$ bin, the statistical uncertainty in the data yield was given by 
\begin{equation}
  \sigma^2_{data} = \mathcal{Y}+ 
  \left(\sum\limits_i^{N_{data}} \sigma_{Qi}\right)^2.
\end{equation}
The relative statistical uncertainty in the acceptance calculation was computed using the expression (see Sec.~5.1 in Ref.~\cite{bevington})
\begin{equation}
\delta \eta/\eta =  \sqrt{\frac{1/ \eta - 1}{N_{\mbox{\scriptsize{raw}}}}}.
\end{equation}

Given the overall azimuthal symmetry of the detector about the beam direction, the data yields in each of the six sectors in the CLAS detector (as tagged by the sector in which the $K^+$ belongs) should have been statistically comparable after acceptance corrections. By examining deviations from this symmetry, we estimated the relative systematic uncertainty in our acceptance calculation to be between 4~and~6$\%$, depending on $\sqrt{s}$. Data collection for the present experiment occurred in bunches of about 10~million event triggers (called ``runs''). Our estimated photon flux normalization uncertainty from a run-wise comparison of the flux-normalized $K^+ \Sigma^0$ yields was $3.2\%$~\cite{my-thesis}. 

In photoproduction experiments, overall normalization uncertainties are often estimated by comparing the total $\pi N$ cross sections with other world data. Since the event trigger for the current experiment required detection of at least two charged tracks, the $\pi N$ channel was not available here. However, a careful study of the cross sections for three different reactions ($\omega p$, $K^+ \Lambda$ and $\eta p$) using the same (present) data set in comparison with other experiments gave a flux normalization uncertainty of $7\%$. Combining this in quadrature with the uncertainty in the run-by-run flux-normalized yield and contributions from photon transmission efficiency ($0.5\%$), live-time ($3\%$) and target density and length ($0.2\%$), we quote an overall normalization uncertainty of $8.3\%$. The last contribution comes from the $\Lambda \to p \pi^-$ branching fraction ($0.5\%$). A list of all the systematic uncertainties pertaining to \dsigma measurements for each of the two topologies is given in Table~\ref{table:systematics}.

\begin{table}
  \centering
  \begin{tabular}{|l|r|r|} \hline
    \multirow{2}{*}{Source of Uncertainty} & \multicolumn{2}{|c|}{ Topology} \\
    \cline{2-3}
    &  $ K^+ p \pi^- (\gamma)$  &  $K^+ p (\pi^-, \gamma) $ \\ \hline
    Particle ID & 0.62\% & 1.8\% \\
    Kinematic Fitter & 3\% & -- \\
    Detector Acceptance & 4\%-6\% & 4\%-6\% \\
    Flux Normalization & 7.7\% & 7.7\% \\
    Detector Live-time & 3\% & 3\% \\
    Transmission efficiency & 0.5\% & 0.5\% \\
    Target Characteristics & 0.2\% & 0.2\%  \\
    $\Lambda \rightarrow p \pi^{-}$ Branching Fraction & 0.5\%  & 0.5\%  \\ \hline
    Overall estimate & 9.7\%-10.7\% & 9.4\%-10.4\%  \\
    \hline
    \end{tabular}
  \caption[]{\label{table:systematics} List of systematic uncertainties for this analysis. The three-track topology has a lower PID uncertainty than the two-track topology but acquires an additional uncertainty from the kinematic fitting systematics.}
\end{table}


\section{\label{section:pol}Recoil Polarization Extraction}

\subsection{``PWA'' method}
The expansion of the production amplitude using partial-wave analysis (PWA) techniques in Sec.~\ref{section:acc} allows for an elegant way of extracting polarization observables for the three-track topology. We first form the two-component wavefunction
\begin{equation}
|\psi_{m_{i}m_{\gamma}}\rangle = \left(\begin{array} {c}
A_{+m_{i}m_{\gamma}}\\
A_{-m_{i}m_{\gamma}}
\end{array} \right)
\label{eqn:pol_psi}
\end{equation}
out of $A_{m_{\Sigma}m_{i}m_{\gamma}}$, suppressing the rest of the indices for the moment. The density matrix is given by
\begin{equation}
\mathbf{\rho} = \sum_{m_{i}m_{\gamma}} |\psi_{m_{i}m_{\gamma}}\rangle \langle\psi_{m_{i}m_{\gamma}}|,
\label{eqn:pol_rho}
\end{equation}
from which the expectation value of the $\Sigma^0$ spin in the direction normal to the production plane (conventionally denoted by $\hat{y}$) is obtained as
\begin{eqnarray}
P_{\Sigma} &=& \frac{\mbox{Tr}\left[\mathbf{\rho}\,\sigma_{y}\right]}{\mbox{Tr}\left[\mathbf{\rho}\,\right]} \nonumber \\
&=& \frac{\displaystyle\sum_{m_{i}m_{\gamma}}  (i A_{+m_{i}m_{\gamma}}A_{-m_{i}m_{\gamma}}^{\ast} - i A_{-m_{i}m_{\gamma}}A_{+m_{i}m_{\gamma}}^{\ast})}{\displaystyle  \sum_{m_{i}m_{\gamma}}  (A_{+m_{i}m_{\gamma}}A_{+m_{i}m_{\gamma}}^{\ast} + A_{-m_{i}m_{\gamma}}A_{-m_{i}m_{\gamma}}^{\ast})}. \nonumber \\
\label{eqn:pol_p_sig_amps}
\end{eqnarray}
The $m_f$ and $m_{\gamma f}$ indices occur only in the $\Sigma^0$ decay part of the above amplitudes as a constant factor that cancels between the numerator and the denominator. The $\Sigma^0$ decay portion of the full amplitude in Eq.~\ref{scat_exp} can therefore be suppressed for $P_{\Sigma}$ extraction.

\subsection{``Traditional'' Method}

\begin{figure*}
  \centering
\subfigure[]{
{\includegraphics[width=2.5in]{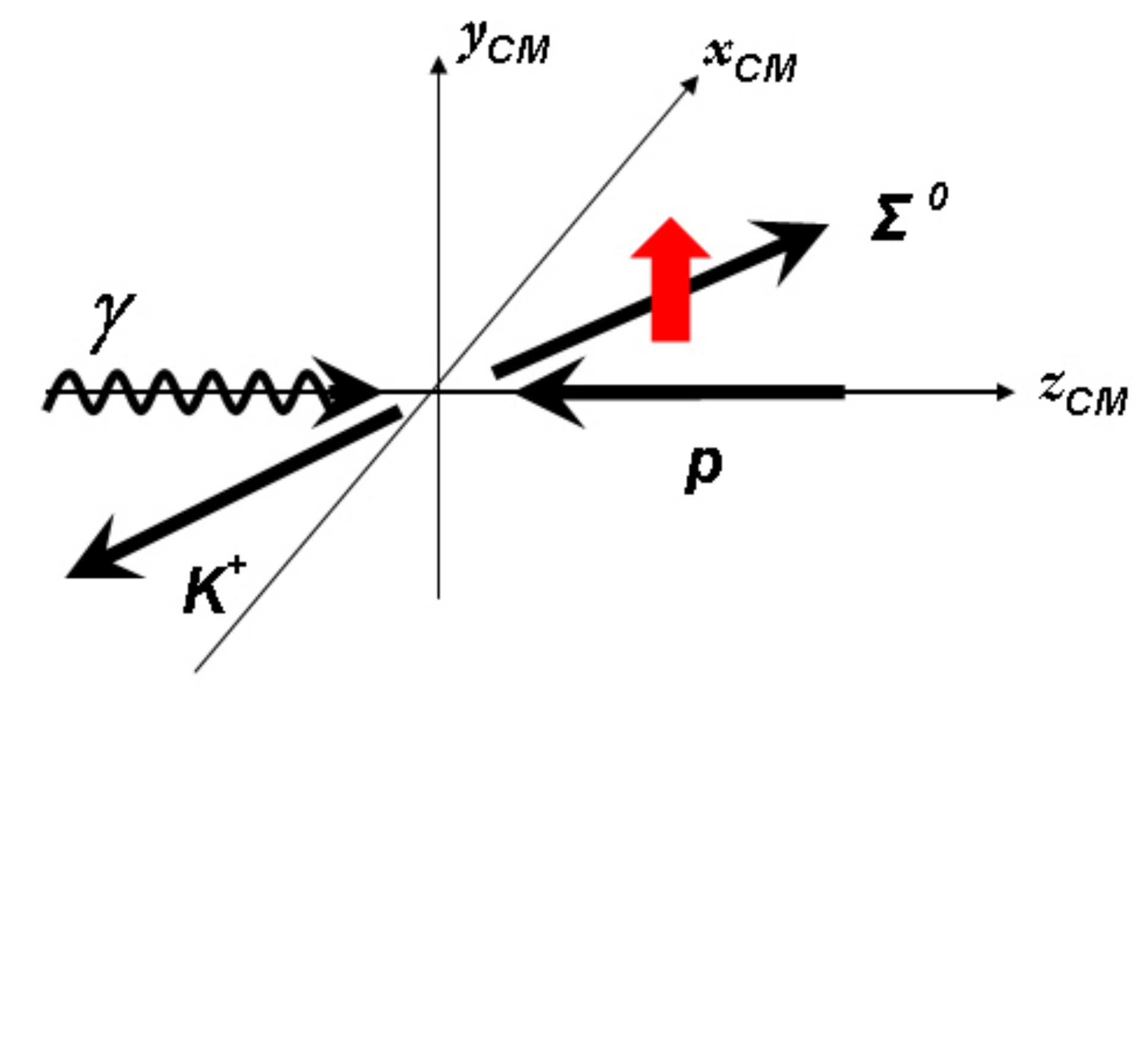}}
}
\subfigure[]{
{\includegraphics[width=2in]{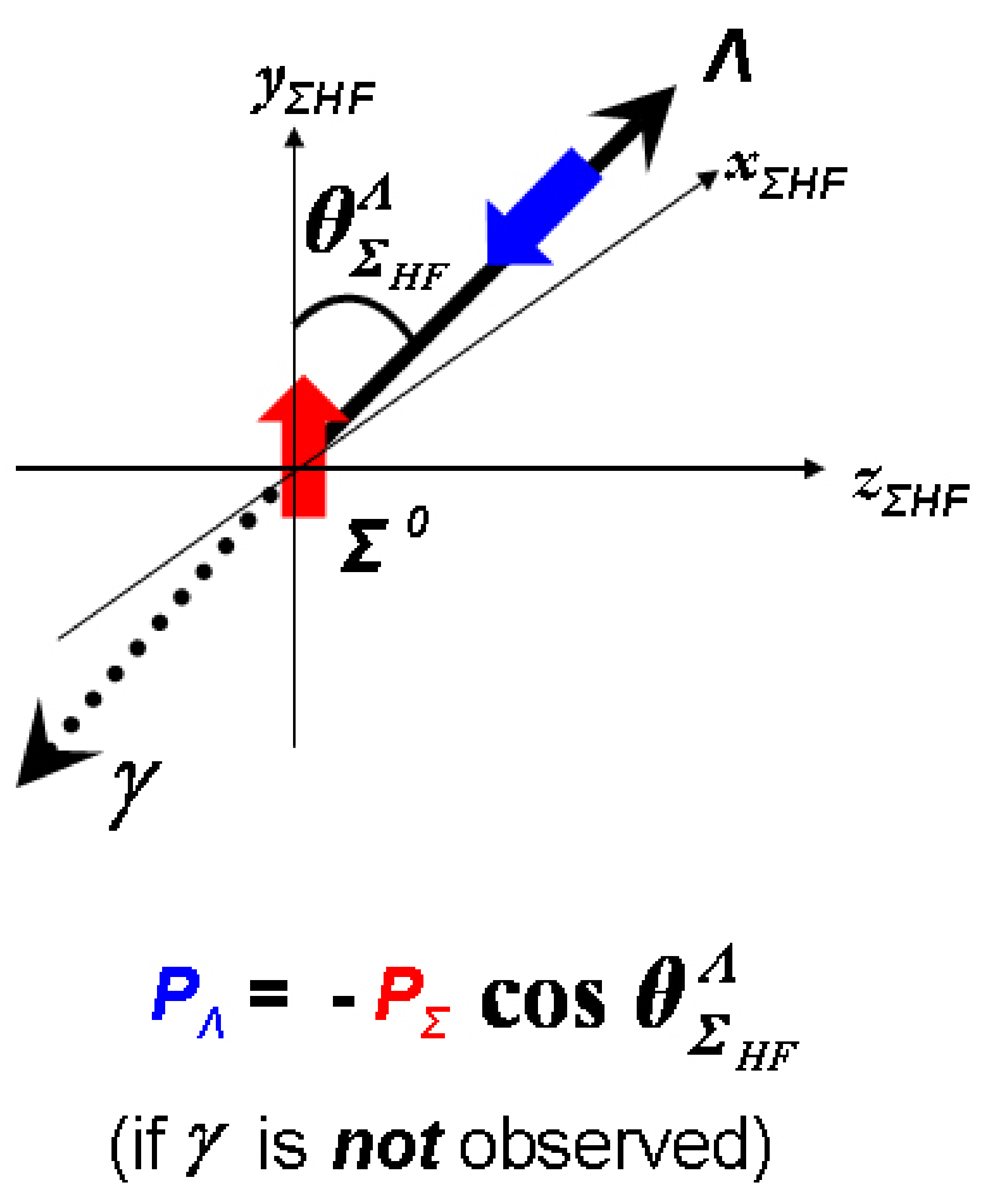}}
}
\subfigure[]{
{\includegraphics[width=2in]{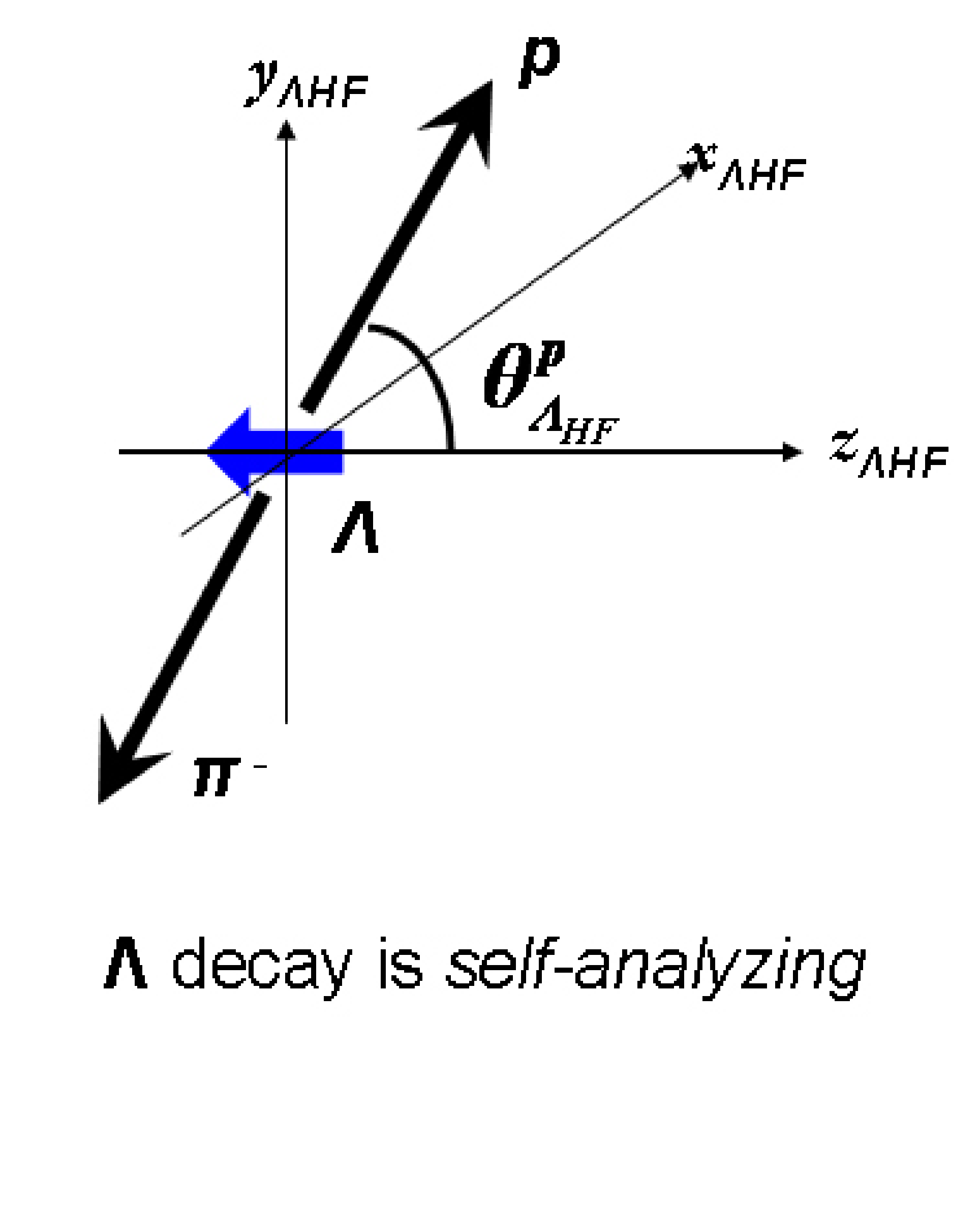}}
}
\caption[]{(Color online) A pictorial representation of the helicity angles $\theta^\Lambda_{\Sigma_{HF}}$ and $\theta^p_{\Lambda_{HF}}$ and the polarization transfer between the $\Sigma^0$ and the $\Lambda$. (a) Shows the $\gamma p \to K^+\Sigma^0$ reaction in the c.m. frame with the $y$-axis as the normal to the production plane. The only component of the induced $\Sigma^0$ spin measurable in the current experiment is along the normal to the production plane, shown by the bold arrow in red. (b) Shows the $\Sigma^0 \to \Lambda \gamma$ decay in the $\Sigma^0$ helicity frame and (c) shows the $\Lambda \to p \pi^-$ decay in the $\Lambda$ helicity frame. See text for details.}
 \label{fig:pol_helicityframe}
\end{figure*}

We first define what we mean by the ``helicity frame'' of a particle. The helicity frame (HF) of any particle is given by an initial rotation that aligns its direction of motion along the $z$-axis, followed by a subsequent boost to its rest frame. $\theta^p_{\Lambda HF}$ is then defined as the angle between the proton and the $\Lambda$ momenta, as measured in the $\Lambda$ helicity frame, while $\theta^{\Lambda}_{\Sigma HF}$ is the angle between the normal to the production plane (assumed to be the $y$-axis) and the $\Lambda$ direction, as measured in the $\Sigma^0$ helicity frame. A pictorial description of these two angles is given in Fig.~\ref{fig:pol_helicityframe}. 

Fig.~\ref{fig:pol_helicityframe}a shows $\gamma p \to K^+\Sigma^0$ reaction in the c.m. frame, where the $z$-axis is along the beam direction and the $y$-axis is normal to the production plane. As mentioned in the introduction, if both the beam and target are unpolarized, as in the current experiment, parity considerations imply that the induced $\Sigma^0$ polarization can only be along the normal to the production plane. This is shown by the bold arrow in red in Fig.~\ref{fig:pol_helicityframe}a. 

To go the $\Sigma^0$ helicity frame from the c.m. frame, we first rotate our system so that the $\Sigma^0$ momentum points along the $z$-axis and then perform a boost to the $\Sigma^0$ rest frame. Fig.~\ref{fig:pol_helicityframe}b shows the $\Sigma^0 \to \Lambda \gamma$ decay in the $\Sigma^0$ helicity frame. Since the outgoing photon (shown by the dotted arrow) was not detected in our experiment, the polarization transfer from the $\Sigma^0$ to the $\Lambda$ is given by (see Ref.~\cite{my-thesis} for a derivation)
\begin{equation}
 P_\Lambda = - P_\Sigma \cos\theta^{\Lambda}_{\Sigma HF}.
\label{eqn:pol_sigma_decay}
\end{equation}
Note that in terms of spin structure, the $\Sigma^0 \rightarrow \Lambda \gamma$ reaction is a {\bf $\frac{1}{2}\rightarrow\frac{1}{2}\oplus1$} decay, while Eq.~\ref{eqn:pol_sigma_decay} is obtained after averaging over the spin projections of the unobserved outgoing photon. Thus, there is a step of ``dilution'' in the accessible $\Sigma^0$ spin information that occurs here.

In the next step, we go to the $\Lambda$ helicity frame from the $\Sigma^0$ helicity frame by making a rotation that aligns the $z$-axis with the $\Lambda$ direction, followed by a boost to the $\Lambda$ rest frame. The $\Lambda \to p \pi^-$ decay (see Fig.~\ref{fig:pol_helicityframe}c) is a {\em self-analyzing} reaction. That is, the $\Lambda$ polarization information is contained in the intensity distribution as
\begin{equation}
\mathcal{I}  \sim (1 + \alpha P_{\Lambda}\cos{\theta^p_{\Lambda HF}}) 
\label{eqn:pol_lambda_decay},
\end{equation}
where $\alpha = 0.642 \pm 0.013$ is the $\Lambda$ weak decay asymmetry~\cite{pdg}. Combining Eqs.~\ref{eqn:pol_sigma_decay} and~\ref{eqn:pol_lambda_decay}, the final intensity distribution is given as
\begin{equation}
\mathcal{I}  \sim (1 - \alpha P_{\Sigma} \cos{\theta^{\Lambda}_{\Sigma HF}}  \cos{\theta^{p}_{\Lambda HF}}).
\label{eqn:pol_traditional}
\end{equation}
Traditionally, the extraction of $P_\Sigma$ has been made using this intensity distribution.

In addition to the ``dilution'' mentioned earlier, a further step of ``dilution'' occurs if one does not have access to the $\Lambda$ momentum. It can be shown (see Appendix A in Ref.~\cite{bradford-cxcz}) that if the $\Sigma^0$-$\Lambda$ spin-transfer information is averaged over, then Eq.~\ref{eqn:pol_traditional} is replaced by 
\begin{equation}
\mathcal{I}  \sim (1 - \nu \alpha P_{\Sigma} \cos{\theta^p_{\Sigma HF}}),
\label{eqn:pol_traditional_2track}
\end{equation}
where $\theta^p_{\Sigma HF}$ is the angle between the outgoing proton's momentum and the normal to the $K^+ \Sigma^0$ production plane as measured in the $\Sigma^0$ helicity frame and $\nu \approx \frac{1}{3.90}$ is a ``dilution factor''. Since the $\pi^-$ from the $\Lambda$ decay was not detected in the two-track topology, the $\Lambda$ momentum could not be reconstructed. Therefore, Eq.~\ref{eqn:pol_traditional_2track} applies instead of Eq.~\ref{eqn:pol_traditional} for the two-track topology.


\section{\label{section:results}Results}

\subsection{\label{section:results:dsigma}Differential cross sections}

\begin{figure*}[p]
  \centering
\includegraphics[width=7in]{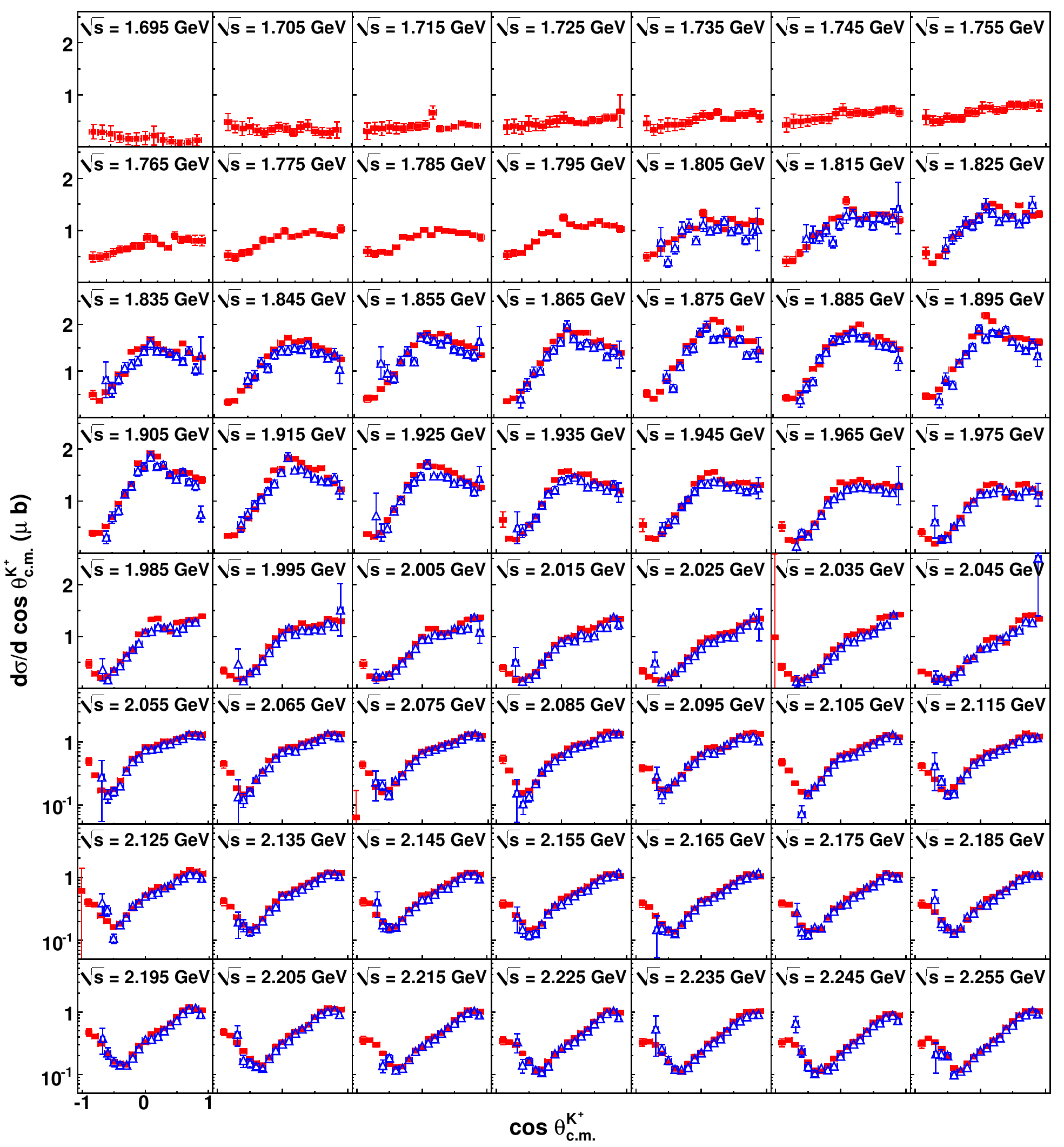}
\caption[]{\label{fig:dsig_2tr_3tr0}
  (Color online) 
  $\frac{d\sigma}{d\cos \theta_{\mbox{\scriptsize{c.m.}}}^{K^+}}$ ($\mu$b) {\em vs.} \cmangle: Differential cross section results for the two topologies in the energy range 1.69~GeV~$\leq \sqrt{s} <$~2.26~GeV. The centroid of each 10-MeV-wide bin is printed on the plots. Results from the two-track analysis are represented by red squares, and those from the three-track analysis by blue triangles. Note that we do not present results in the $\sqrt{s} = 1.955$~GeV bin (see Sec.~\ref{section:norm}) and that the $y$-axes are set to logarithmic scales from $\sqrt{s} = 2.055$~GeV onwards. All error bars represent statistical uncertainties only. 
}
\end{figure*}

\begin{figure*}[p]
  \centering
\includegraphics[width=7in]{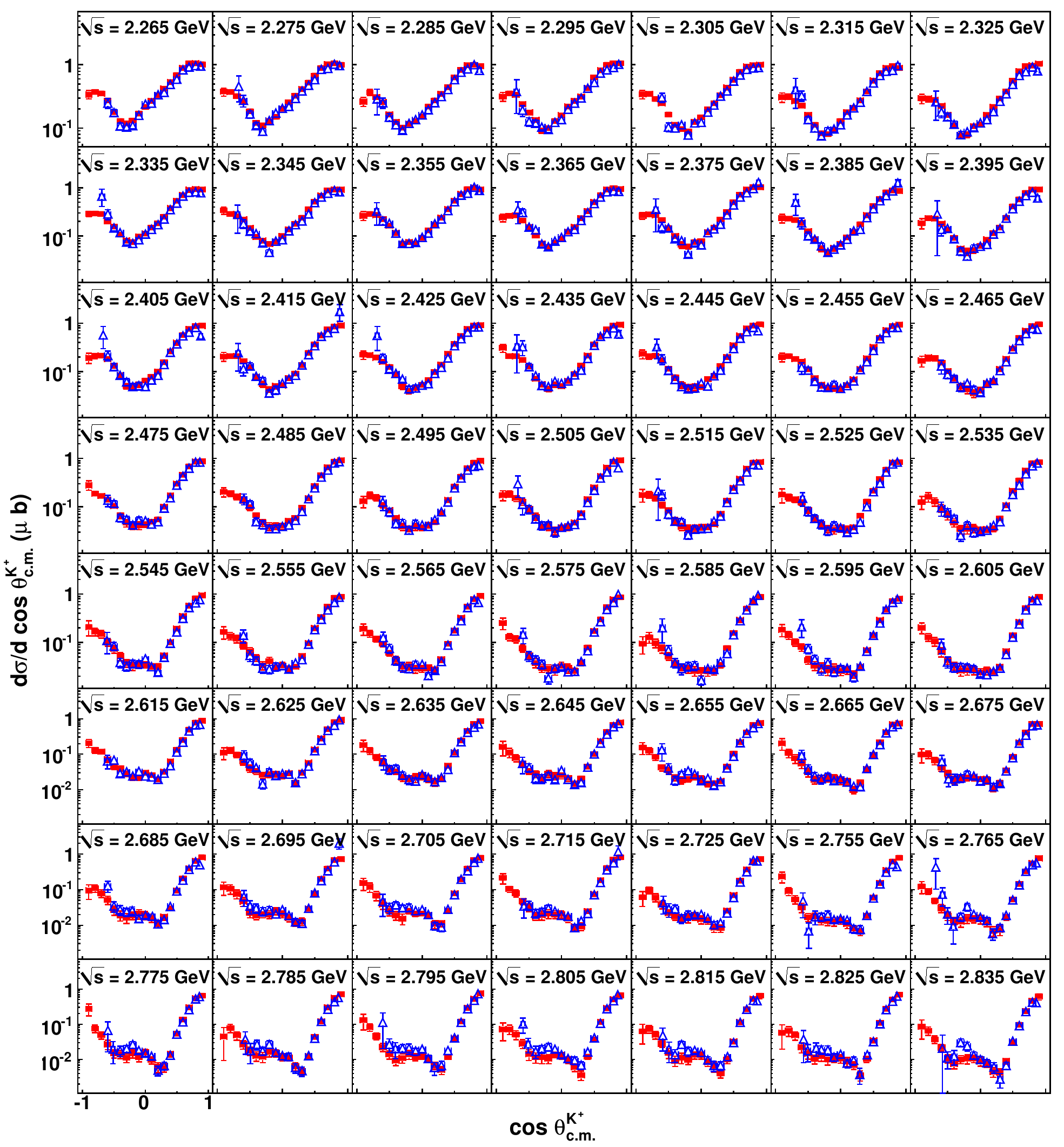}
\caption[]{\label{fig:dsig_2tr_3tr1}
  (Color online) 
  $\frac{d\sigma}{d\cos \theta_{\mbox{\scriptsize{c.m.}}}^{K^+}}$ ($\mu$b) {\em vs.} \cmangle: Differential cross section results for the two topologies in the energy range 2.26~GeV~$\leq \sqrt{s} <$~2.84~GeV. The centroid of each 10-MeV-wide bin is printed on the plots. Results from the two-track analysis are represented by red squares, and those from the three-track analysis by blue triangles. Note that we do not present results in the $\sqrt{s} = 2.735$ and 2.745~GeV bins (see Sec.~\ref{section:norm}) and that the $y$-axes are set to logarithmic scales everywhere. All error bars represent statistical uncertainties only. 
}
\end{figure*}

\begin{figure*}
  \centering
\includegraphics[width=7in]{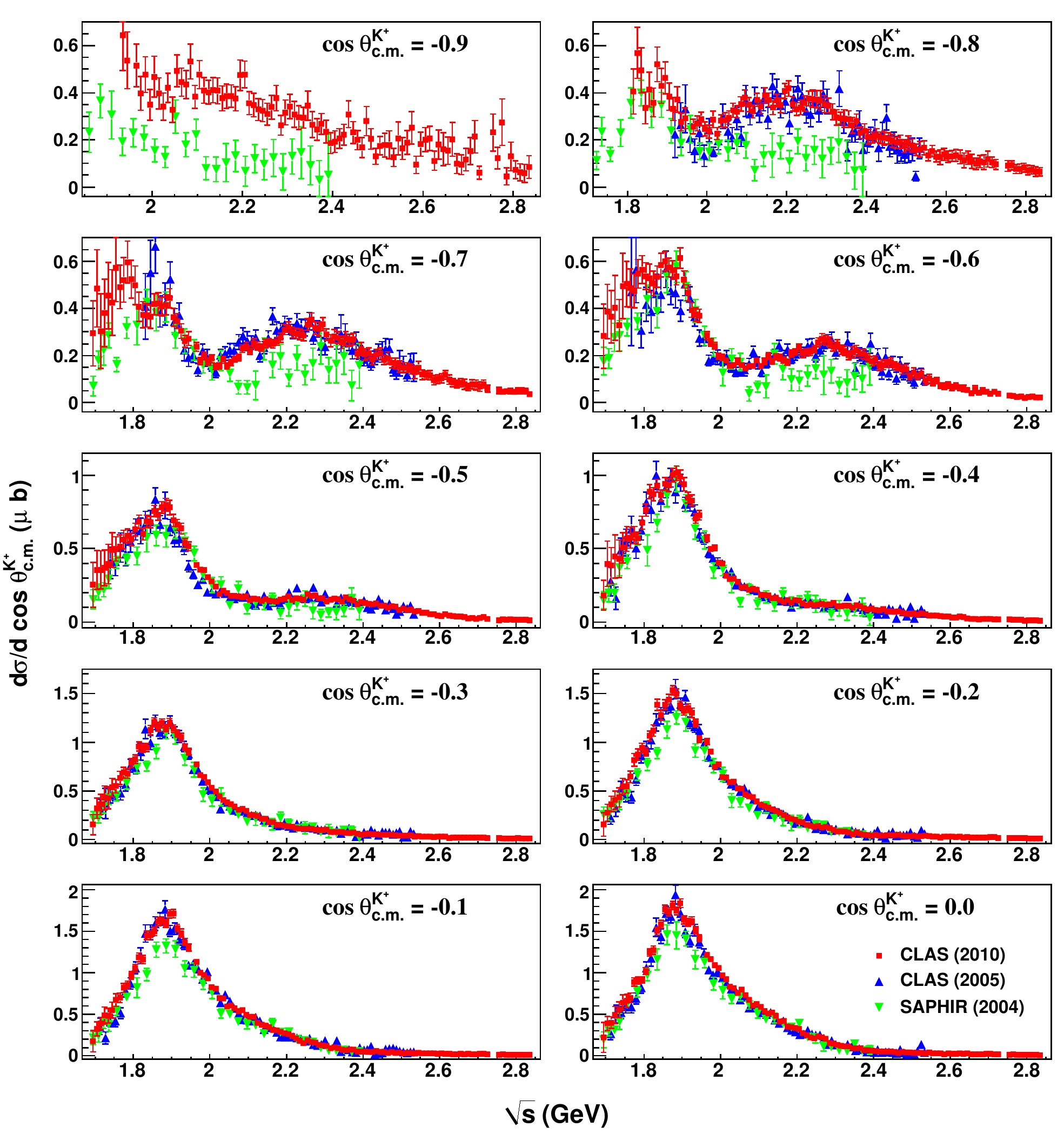}
\caption[]{\label{fig:dsig_world_data0}
  (Color online) 
  $\frac{d\sigma}{d\cos \theta_{\mbox{\scriptsize{c.m.}}}^{K^+}}$ ($\mu$b) {\em vs.} $\sqrt{s}$ in the backward-angles: final CLAS (present analysis) differential cross section results as the weighted average of the two topologies are in red squares. Previous CLAS results~\cite{bradford-dcs} are in blue up-triangles while green down-triangles are results from SAPHIR~\cite{glander-saphir}. All error bars represent statistical uncertainties only.
}
\end{figure*}

\begin{figure*}
  \centering
\includegraphics[width=7in]{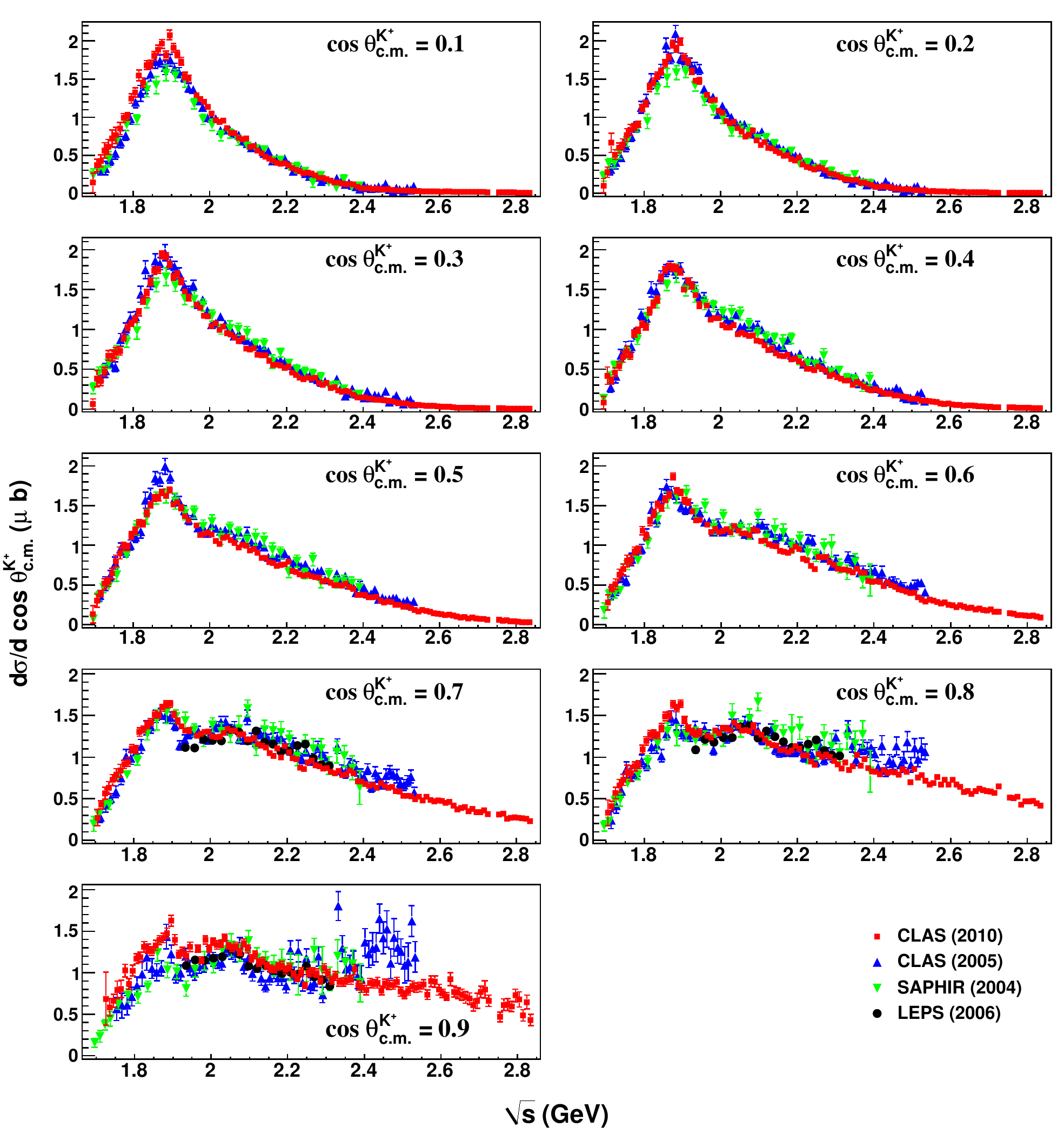}
\caption[]{\label{fig:dsig_world_data1}
  (Color online) 
  $\frac{d\sigma}{\cos \theta_{\mbox{\scriptsize{c.m.}}}^{K^+}}$ ($\mu$b) {\em vs.} $\sqrt{s}$ in the forward-angles: final CLAS (present analysis) differential cross section results as the weighted average of the two topologies are in red squares. Previous CLAS results~\cite{bradford-dcs} are in blue up-triangles while green down-triangles are results from SAPHIR~\cite{glander-saphir}. The black circles represent LEPS measurements~\cite{kohri-leps} in the forward-most angular bins. All error bars represent statistical uncertainties only.
}
\end{figure*}

Figs.~\ref{fig:dsig_2tr_3tr0}~and~\ref{fig:dsig_2tr_3tr1} show our differential cross section results separately for the two topologies. The binning in $\sqrt{s}$ is 10~MeV and 0.1 in \cmangle for both cases. The three-track results span from 1.80~GeV~to~2.84~GeV in $\sqrt{s}$, while the two-track results start closer to the threshold at $\sqrt{s}=$~1.69~GeV~to~2.84~GeV. The higher acceptance also allows greater coverage of the backward-angles for the two-track analysis. 

Given the widely different analysis techniques employed for the two topologies, the agreement between the two results is significant. The three-track analysis made use of a kinematic fitter and a PWA fit-based, weighted acceptance calculation method, neither of which was available for the two-track analysis. The slight remaining differences between the two topologies are within the overall 10-11$\%$ systematic uncertainties. 
 
The consistency between results from the two topologies allow us to quote our final differential cross sections as the weighted mean according to 
\begin{equation}\label{eq:weighted_mean}
  \overline{x}(\sqrt{s},\cos \theta_{\mbox{\scriptsize{c.m.}}}^{K^+}) = \frac{x_2 \sigma_3^2 - \rho_{\mbox{\scriptsize{eff}}}\,(x_2 + x_3)\sigma_2 \sigma_3 + x_3 \sigma_2^2}{\sigma_2^2 - 2 \rho_{\mbox{\scriptsize{eff}}}\, \sigma_2 \sigma_3 + \sigma_3^2} ,
\end{equation}
where $x_{2,3}$ are the differential cross sections and $\sigma_{2,3}$ are the associated statistical uncertainties for the two- and three-track results, respectively. Here, $\rho_{\mbox{\scriptsize{eff}}} \approx 0.33$ is an effective degree of correlation that takes into account the ratio between the total signal yields in the two data sets. The statistical uncertainty on the mean value is given by
\begin{equation}\label{eq:weighted_err}
  \overline{\sigma}(\sqrt{s},\cos \theta_{\mbox{\scriptsize{c.m.}}}^{K^+}) = \sqrt{ \frac{ \sigma_2^2 \sigma_3^2 (1 - \rho_{\mbox{\scriptsize{eff}}}^2) }{ \sigma_2^2 - 2 \rho_{\mbox{\scriptsize{eff}}}\,\sigma_2 \sigma_3  + \sigma_3^2}} .
\end{equation}
The derivation of these expressions and the computation of $\rho_{\mbox{\scriptsize{eff}}}$ can be found in Ref.~\cite{my-thesis}.

Figs.~\ref{fig:dsig_world_data0}~and~\ref{fig:dsig_world_data1} show the final differential cross sections for the present experiment, presented at 2089 kinematic points, in comparison with previously published high statistics measurements. The latter consists of results from a previous CLAS experiment by Bradford {\em et al.} 2005~\cite{bradford-dcs}, a SAPHIR analysis by Glander {\em et al.} 2004~\cite{glander-saphir} and a set of more recent forward-angle measurements using the LEPS detector by Kohri \emph{et al.} 2006~\cite{kohri-leps}. Overall, there is good consistency among the different data sets. There is a peak at $\sqrt{s}\approx 1.9$~GeV prominent over the entire angular range. In the forward-angle bins ($\cos \theta_{CM}^{K^+} \gtrsim 0.7$), there seems to be an initial dip in the cross section just after the 1.9~GeV peak and a subsequent rise, indicative of a smaller second peak around $\sqrt{s}\approx 2.1$~GeV. This feature was also pointed out in the previous CLAS analysis~\cite{bradford-dcs} and is present in the latest LEPS data~\cite{kohri-leps} as well.

Some notable localized discrepancies also occur between the different results. Chiefly, this pertains to the ``hump'' at backward-angles at $\sim2.2$~GeV seen in the previous CLAS results~\cite{bradford-dcs}, but not prominent in the SAPHIR data~\cite{glander-saphir}. The present CLAS results, however, clearly confirm this structure. In fact, the SAPHIR differential cross sections seem to be generally lower (or ``flatten out'') towards the backward-angles, as compared to CLAS results, for both $K^+ \Sigma^0$ and $K^+ \Lambda$~\cite{klam_prc}. Generally speaking, for both hyperons, the two CLAS results are in very good agreement. Recent LEPS data for $K^+ \Lambda$ in the backward-angles~\cite{klam_leps_hicks} also shows good agreement with the latest CLAS $K^+ \Lambda$ data~\cite{klam_prc}. So it is possible that the flattening out at backward-angles is due to some overall normalization issue in the SAPHIR data. We also note that in the two forward-most angular bins, the previous CLAS results~\cite{bradford-dcs} showed an unphysical rise in the differential cross sections above $\sqrt{s}\approx 2.2$~GeV, that was attributed to systematic uncertainties in the acceptance (systematic uncertainties are not included in Figs.~\ref{fig:dsig_world_data0}~and~\ref{fig:dsig_world_data1}). The new results do not show this unphysical rise. 

\subsection{\label{section:results:p_sigma}Recoil polarizations}

As explained earlier in Sec.~\ref{section:pol}B, going from Eq.~\ref{eqn:pol_traditional} (three-track) to Eq.~\ref{eqn:pol_traditional_2track} (two-track) represents a second step of dilution in the polarization extraction procedure. This arises from the fact that the polarization transfer between $\Lambda$ and $\Sigma^0$ in the decay $\Sigma^0 \to \Lambda \gamma$ remains unknown for the two-track topology. The effect of this dilution was studied~\cite{my-thesis} by making use of the traditional method of extraction for the three-track topology and comparing the results from using Eq.~\ref{eqn:pol_traditional} to that from Eq.~\ref{eqn:pol_traditional_2track}. On the average, the two results agreed very well, but the scatter in the polarization data from the diluted expression (Eq.~\ref{eqn:pol_traditional_2track}) was much larger. Quantitatively, by comparing the point-by-point ratios of the error bars, the uncertainties in the polarization extracted via Eq.~\ref{eqn:pol_traditional_2track} were found to be about 2.8 times larger than those via Eq.~\ref{eqn:pol_traditional}.  

To avoid this extra step of dilution inherent in the two-track topology, we have chosen to present $P_{\Sigma}$ results using the three-track data set wherever this is possible. This covers the energy range 1.80~GeV~$\leq\sqrt{s}\leq$~2.84~GeV and the angular range -0.55~$\leq\cos \theta_{\mbox{\scriptsize{c.m.}}}^{K^+}\leq$~0.95. Our \cmangle binning is 0.1, the same as for the differential cross sections. However, to bolster statistics, we bin wider in $\sqrt{s}$ (minimum 30-MeV-wide bins) and demand a minimum occupancy ($Q$-value weighted yields from Sec.~\ref{section:sig_bkgd}) of 200 events for every kinematic point at which a measurement is reported.

\begin{figure}
\begin{center}
\includegraphics[width=0.45\textwidth]{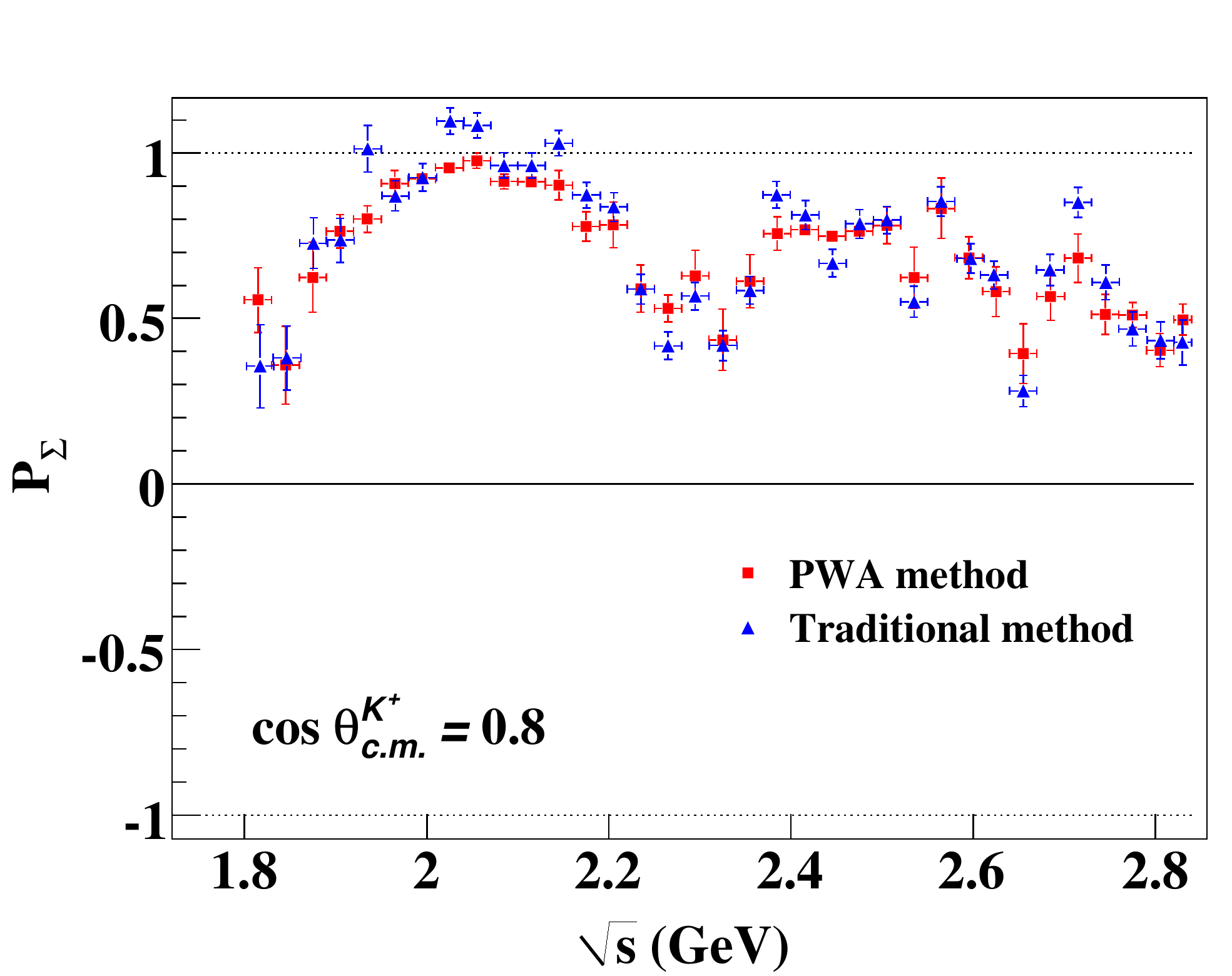}
\caption[]{\label{fig:pol_pwa_intensity} (Color online) Comparison of the three-track $P_\Sigma$ obtained from the partial-wave expansion approach of Eq.~\ref{eqn:pol_p_sig_amps} (red squares) and the more traditional intensity fit based approach of Eq.~\ref{eqn:pol_traditional} (blue triangles). The agreement is excellent. At places where $P_\Sigma$ approaches unity, the intensity method polarizations tend to overshoot slightly. The uncertainties for the PWA method are calculated indirectly from the statistical spread in the data, while the traditional method uncertainties directly come from the covariance matrix of a fit.}
\end{center}
\end{figure}

From our discussion in Sec.~\ref{section:pol}, recoil polarizations for the three-track topology can be extracted either by the intensity distribution expression, or by a more sophisticated partial-wave expansion method. Fig.~\ref{fig:pol_pwa_intensity} shows a comparison between the two methods. The overall agreement is excellent, emphasizing the fact that the underlying physics extracted in both approaches was the same. Note that polarizations from the ``PWA'' method are constrained by Eq.~\ref{eqn:pol_p_sig_amps} to lie within the physical limits of $\pm1.0$, while the ``Traditional'' method polarization results are not constrained in any such fashion. We found that in some kinematic regions where the degree of induced $\Sigma^0$ polarization was sufficiently high, the ``Traditional'' method sometimes gave $P_\Sigma$ values that were greater than unity. In all such cases, the ``PWA'' method gave $P_\Sigma$ values that were close to but always smaller than unity, demonstrating the consistency among the two approaches. 

The large number of parameters in the PWA fit led to certain difficulties in estimating the $P_\Sigma$ uncertainties from the ``PWA'' method. Given the large number of waves employed in our PWA fits, sometimes the waves interfered strongly amongst each other with only a small contribution to the final result. However, the covariance matrix elements corresponding to these waves were typically non-zero and the accumulated contribution from such small but non-zero covariance matrix elements gave rise to unphysically large estimated uncertainties. The uncertainties for the ``PWA'' method have therefore been obtained from the statistical spread between neighboring $\sqrt{s}$ measurements for a given \cmangle bin. A similar approach was also followed in two other previous CLAS analyses of the $\omega p$~\cite{omega_prc} and $K^+\Lambda$~\cite{klam_prc} photoproduction channels and found to give reasonable results.

The estimated uncertainties from the ``Traditional'' method are simply the uncertainties obtained from unbinned maximum likelihood fits to the intensity distributions in Sec.~\ref{section:pol}B. As shown in Fig.~\ref{fig:pol_pwa_intensity} the uncertainties from the two methods are comparable, demonstrating that the likelihood fit uncertainties faithfully represent the statistical spread in the data. In keeping with the internal consistency between the two methods, we report our final three-track recoil polarization measurements as follows. The values of the polarizations are the ones from the ``PWA'' method while the statistical uncertainties are the uncertainties obtained from the ``Traditional'' method. We reiterate the fact here that the PWA expansion in Eq.~\ref{scat_exp} was specifically tuned to represent distributions in all kinematic variables, in particular, the intensity distributions given in Sec.~\ref{section:pol}B. Therefore, the two methods are completely equivalent. The advantage of the ``PWA'' method is that it yields values for $P_\Sigma$ that are within the physical limits, while it is easier to estimate the statistical uncertainties using the ``Traditional'' method. 

Our systematic uncertainty for $P_\Sigma$ was estimated to be $\sim 0.03$ from an examination of the systematic difference between the two methods. Unlike \dsigma, $P_\Sigma$ is bound between $\pm 1$, which is why we quote an absolute uncertainty here, instead of a relative percentage uncertainty.

In the backward-angle and/or the near-threshold bins, where statistics are extremely limited for the three-track data set, the polarizations are presented from the two-track analysis. In general, $P_\Sigma$, being an {\em asymmetry} measurement, is much more sensitive to the statistics than the differential cross sections. This is especially true for the limited-statistics backward-angle/near-threshold bins. Therefore, we make a judicious choice when presenting the two-track polarization results, omitting measurements with unreasonably large error bars. In all, we present 459 individual data points. Finally, we also point out that while re-binning the two-track data set in $\sqrt{s}$, we were careful that $\sqrt{s} = 1.8$~GeV always lay on a bin edge. Since the three-track data sets extends from 1.8~GeV onwards, this ensured that there was no kinematic overlap between results from the two topologies.  

\begin{figure*}
  \centering
 \includegraphics[width=7in]{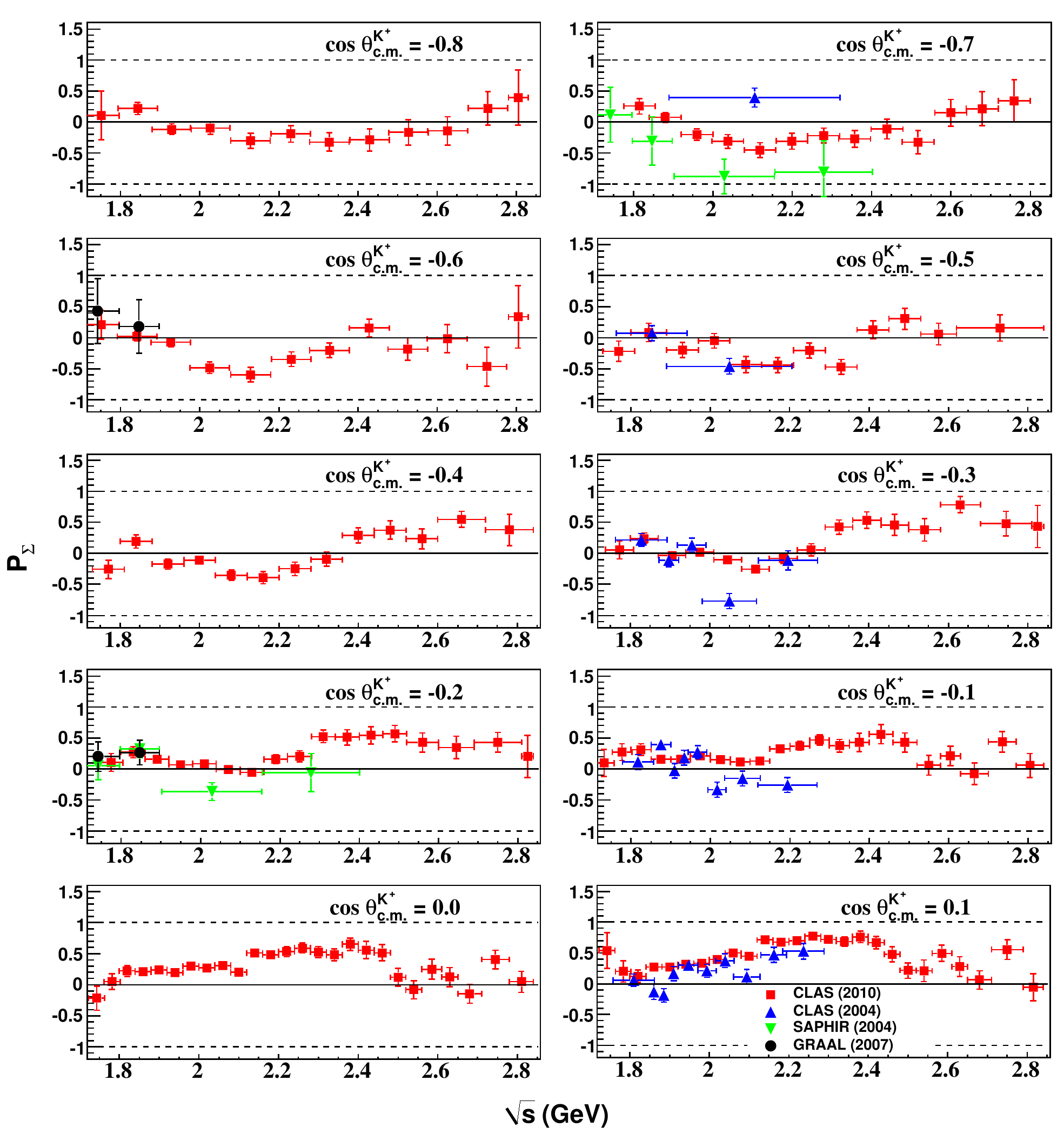}
\caption[]{\label{fig:psig_world_data_1}
  (Color online)
  $P_\Sigma$ {\em vs.} $\sqrt{s}$ : Recoil polarization world data in the backward-angles. CLAS (present analysis) results are in red squares, earlier CLAS~\cite{mcnabb} results in blue up-triangles, SAPHIR~\cite{glander-saphir} in green down-triangles, GRAAL~\cite{lleres-graal} are in black circles. The error bars represent the statistical uncertainties.}
\end{figure*}

\begin{figure*}
  \centering
\includegraphics[width=7in]{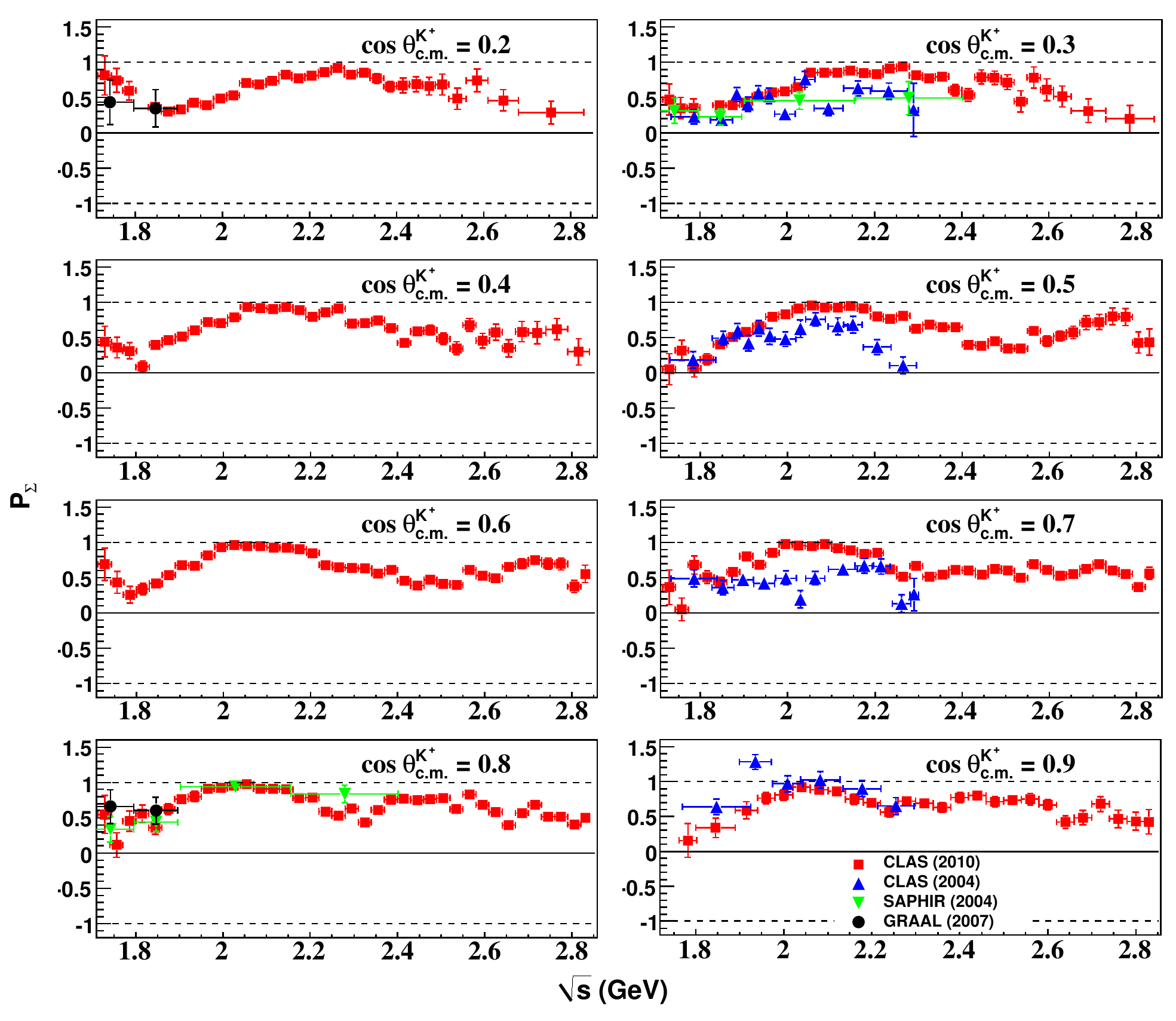}
\caption[]{\label{fig:psig_world_data_2}
  (Color online)
  $P_\Sigma$ {\em vs.} $\sqrt{s}$ : Recoil polarization world data in the forward-angles. CLAS (present analysis) results are in red squares, earlier CLAS~\cite{mcnabb} results in blue up-triangles, SAPHIR~\cite{glander-saphir} in green down-triangles, GRAAL~\cite{lleres-graal} are in black circles. The error bars represent the statistical uncertainties.
}
\end{figure*}

Figs.~\ref{fig:psig_world_data_1} and~\ref{fig:psig_world_data_2} show our final measurements in comparison with some earlier results from CLAS (McNabb \emph{et al.}, 2004~\cite{mcnabb}), SAPHIR (Glander \emph{et al.}, 2004~\cite{glander-saphir}) and GRAAL (Lleres \emph{et al.}, 2007~\cite{lleres-graal}). Previous world data are generally sparse and the contribution from the present analysis brings wide improvements in kinematic coverage and precision. It is noteworthy that all previous measurements of $P_\Sigma$ employed the diluted expression given by Eq.~\ref{eqn:pol_traditional_2track}. Since the majority of our polarization results come from the three-track data set that avoids this dilution, they represent an improvement not just in terms of greater statistical precision, but also in terms of an additional physics precision. Structures that were hinted at by earlier measurements are now mapped out in much greater detail. In the three angular bins $\cos \theta_{\mbox{\scriptsize{c.m.}}}^{K^+} = 0.3$, 0.5 and 0.7, there are a few localized discrepancies between the present results and earlier CLAS measurements~\cite{mcnabb}. We are uncertain of the precise origin of these discrepancies, but given that the present analysis incorporates a vast improvement in statistical precision, makes use of more sophisticated analysis tools and have been checked for internal consistency, the present results supersede the earlier measurements by CLAS.

\begin{figure}
\begin{center}
\includegraphics[width=0.48\textwidth]{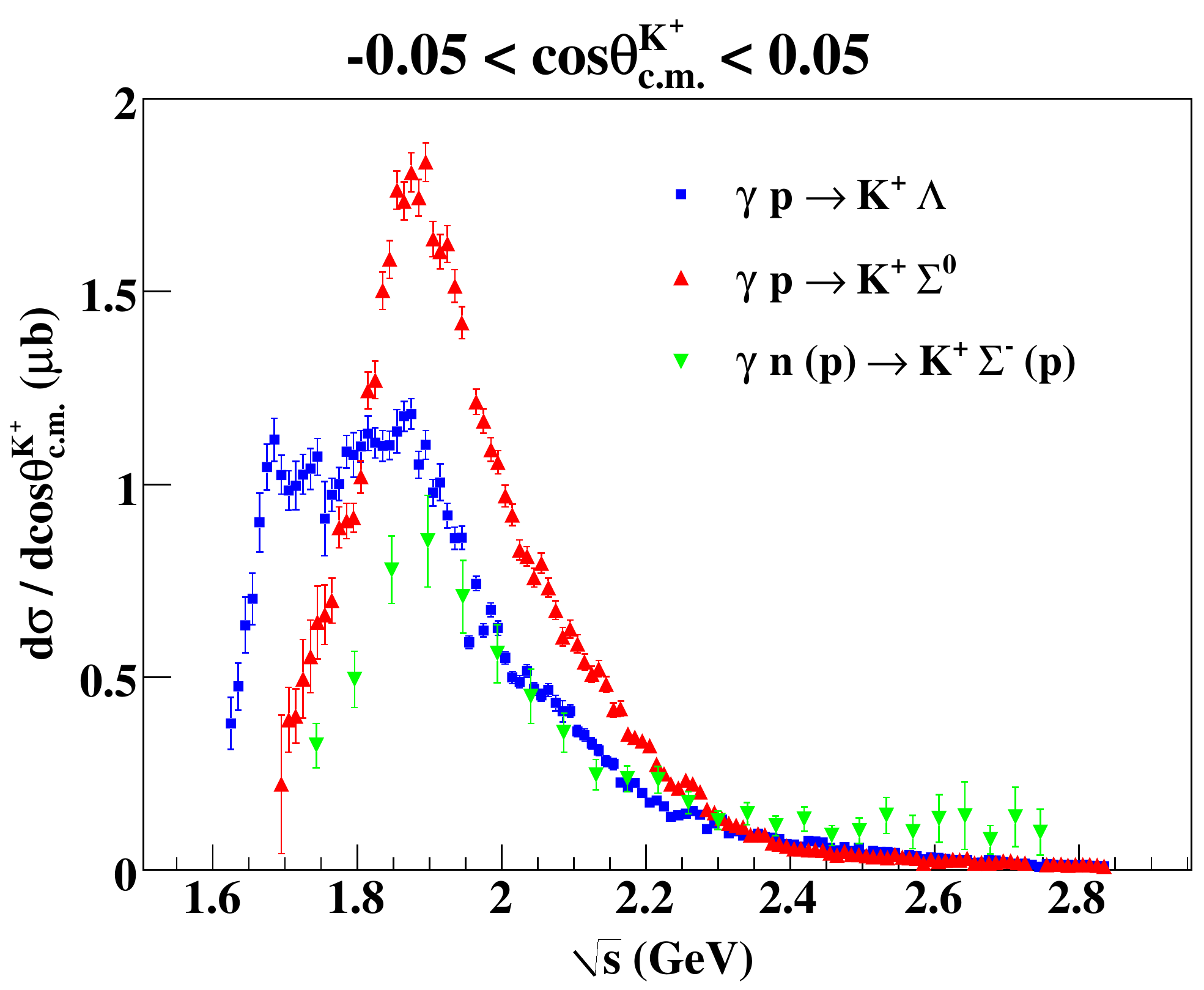}
\caption[]{(Color online) 
Differential cross sections at ${\theta_{\mbox{\scriptsize{c.m.}}}^{K^+} = 90^\circ}$ for $\Lambda$~\cite{klam_prc} (blue squares) and $\Sigma^0$ (red up-triangles) photoproduction from proton and $\Sigma^-$~\cite{sergio} (green down-triangles) photoproduction from deuterium. Except for the extreme forward-angles, all three hyperon channels show a similar peak at $\sqrt{s} \sim 1.9$~GeV.
}
\label{fig:piNscaled}
\end{center}
\end{figure}


\section{ Physics Discussion \label{sect:discussion}}

\subsection{Differential cross sections}

\begin{figure*}
\begin{center}
\subfigure[]{
{\includegraphics[width=3.2in]{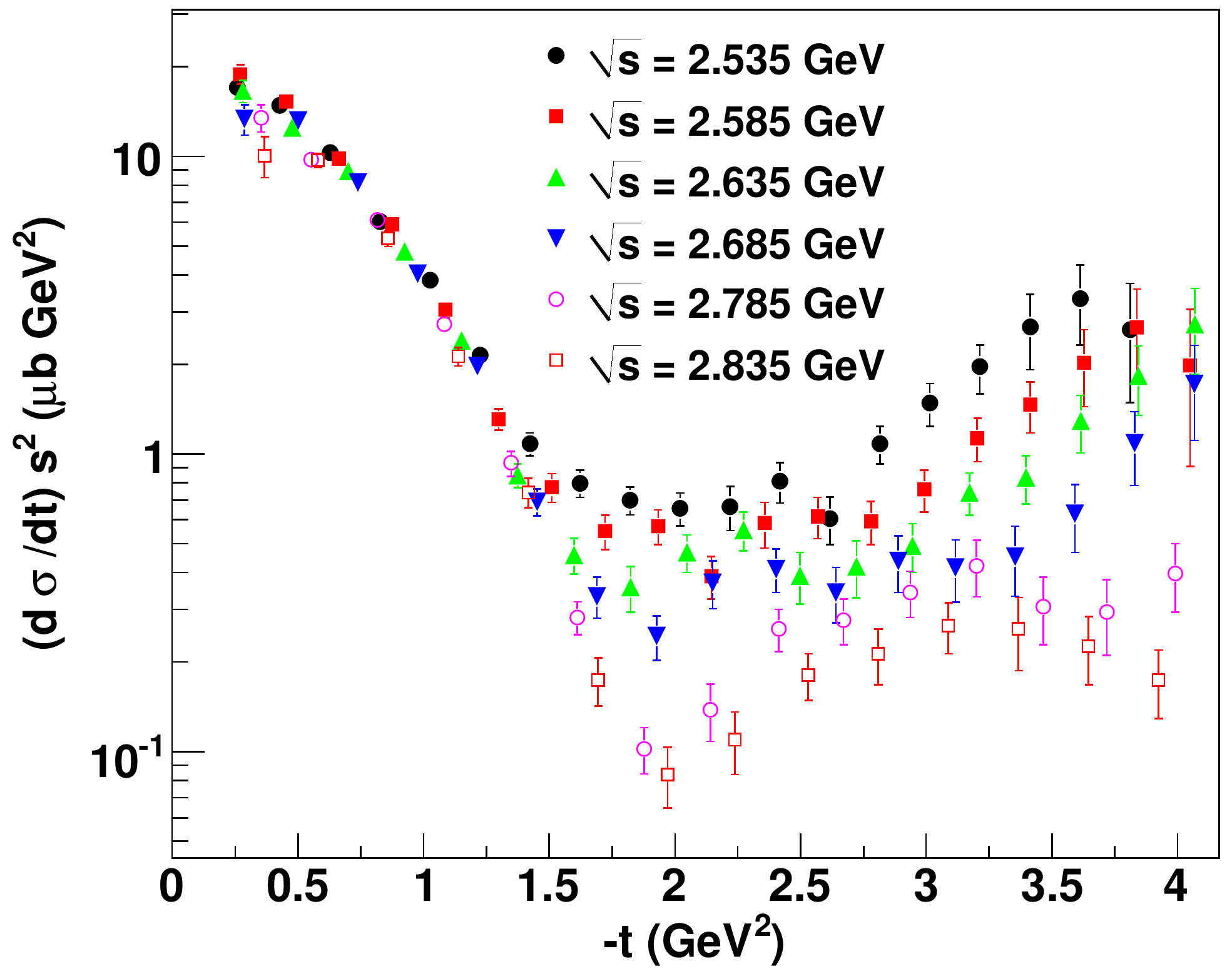}}
}
\hspace{0.7cm}
\subfigure[]{
{\includegraphics[width=3.18in]{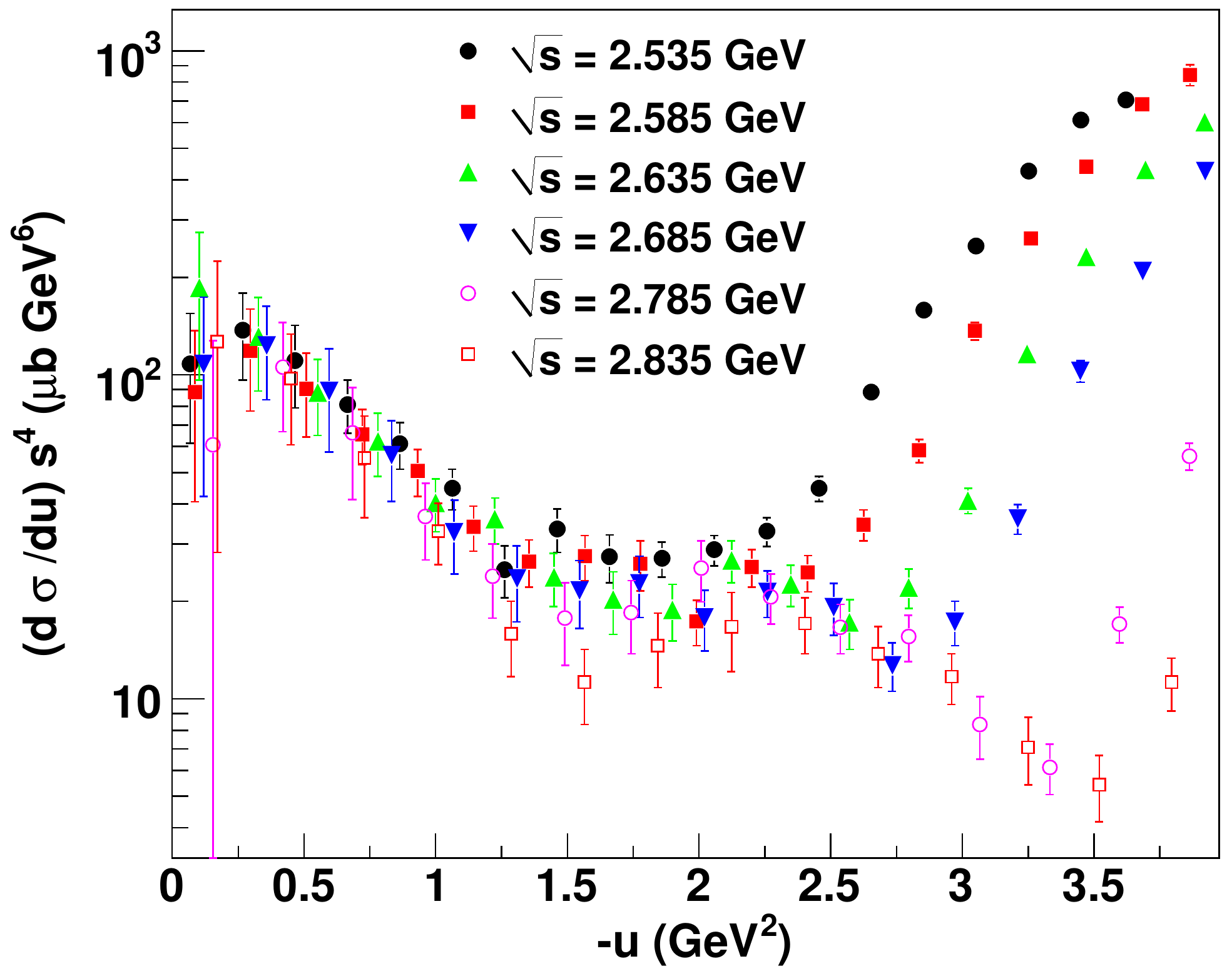}}
}
\caption[]{(Color online)
Regge scaling at high $s$ for: (a) $t$-channel (forward-angle) and (b) $u$-channel (backward-angle) non-resonant processes. The different marker colors and symbols correspond to different $\sqrt{s}$. The scaling behavior appears as the convergent set of points as $t\to 0$ or $u \to 0$.
}
\label{fig:regge_scaling}
\end{center}
\end{figure*}

The most prominent feature in our differential cross section results is the peak at 1900~MeV, visible over the entire angular range. This is not a new feature by itself~\cite{bradford-dcs, glander-saphir}, but its interpretation in terms of $s$-channel baryon resonances has been highly controversial. In the introduction section, we pointed out several possible states, e.g. $D_{13}(1895)$, $S_{31}(1900)$, $P_{31}(1910)$, claimed by different groups. More recently, the Bonn-Gatchina group~\cite{anisovich, bg_nikonov} has claimed that a positive-parity $P_{13}(1900)$ state with a two-star rating in the PDG is able to account for features in both the differential cross section and polarization. There is also a possibility that this structure might not be a ``resonance'' {\em per se}, but simply a strangeness threshold phenomenon. The $\phi$-$N$ bound state mass representing the strangeness threshold is also around 1.9~GeV and it has been shown by Brodsky {\em et al.}~\cite{brodsky_phi_n, gao_phi_prc, qian_plb} that at low energies, via pure gluonic exchanges, QCD can give rise to an attractive Van der Waals force, so that $\phi$ and $N$ can indeed bind. It is interesting to note that recent CLAS $K^+\Sigma^-$ photoproduction data~\cite{sergio} also show a prominent peak at around 1.9~GeV. Fig.~\ref{fig:piNscaled} shows the differential cross sections for the three hyperons $\Lambda$, $\Sigma^0$ and $\Sigma^-$ at $ \theta_{\mbox{\scriptsize{c.m.}}}^{K^+} = 90^\circ$. It was pointed out in a previous CLAS paper~\cite{scaling_piN} that $\gamma N \to \pi N$ scaled differential cross sections ($s^7 \times d \sigma/dt$) at $\theta_{\mbox{\scriptsize{c.m.}}}^\pi=90^\circ$ also show a similar ``rise'' commencing from around 1.9~GeV. It is therefore possible that all these structures are connected to the same universal $s \bar{s}$ production threshold. 

The structure at $\sqrt{s} \approx 2.2$~GeV in the backward-angles is also quite interesting. Recently, the LEPS Collaboration~\cite{leps_eta_backward} has published $\eta p$ photoproduction results at backward-angles that show a bump-like structure in the differential cross sections at around $E_\gamma = 2.0$~GeV. The claim has been that the bump is absent in the $\eta'$, $\omega$ and $\pi^0$ channels, pointing towards the conjecture that it is due to some resonance that couples strongly to the strange sector (the $\eta$ has a higher $s \bar{s}$ component than $\eta'$). Thus, it is possible that the structures seen in the $\eta$ and $\Sigma^0$ channels are related. 

One of the long-standing issues in model calculations has been the contribution from background non-resonant processes. The additional 300-MeV coverage at the higher energy end provided by our results should be very useful in clearing up this issue. At high $\sqrt{s}$, where there are presumably few $s$-channel resonance contributions, $t$-channel processes are known to dominate at forward-angles. In the Regge description, for high $s$ forward-scattering, the production amplitude should scale as $\mathcal{A}(s,t) \sim s^{-\alpha(t)}$, where the Mandelstam variable $t$ denotes the exchange momentum squared and $\alpha(t)$ is the Regge trajectory of the exchanged particle (Reggeon). Since $d \sigma / dt \sim s^{-2} |\mathcal{A}|^2$, $d \sigma / dt$ is expected to scale as $s^{2 (\alpha_{\mbox{\scriptsize{eff}}} -1)}$, where $\alpha_{\mbox{\scriptsize{eff}}}$ is an ``effective'' Regge trajectory for multiple Reggeon exchange. Such an $s^{-2}$ (signifying $\alpha_{\mbox{\scriptsize{eff}}} \approx 0$) scaling was already seen at lower energies ($\sqrt{s} \leq 2.5$~GeV) in the previous CLAS analysis by Bradford {\em et al.}~\cite{bradford-dcs}. A very similar scaling phenomena is seen in the present data at the highest energies, as shown in Fig.~\ref{fig:regge_scaling}a. Following Guidal {\em et al.}~\cite{guidal}, the previous CLAS paper~\cite{bradford-dcs} also pointed out that if one assumes the $t$-channel processes to be dominated by $K^+$ and $K^\ast (892)$ exchanges, the effective Regge trajectory could be simply explained as $\alpha_{\mbox{\scriptsize{eff}}} = \alpha_{K^+} + \alpha_{K^\ast} \approx 0$ for $t \to 0$. 

The Guidal-Laget-Vanderhaeghen paper~\cite{guidal} also pointed out that a similar scaling should be observed at high $s$, $u \to 0$ (backward-angle scattering). However, previous world data did not have enough statistical precision at backward-angles to demonstrate this conclusively. Fig.~\ref{fig:regge_scaling}b shows the $u$-channel scaling behavior in the present analysis, as demonstrated by the strongly collimated peak as $|u|$ decreases. We also found that an $s^{-4}$ scaling worked better than $s^{-2}$ here. Since the Reggeons involved in $u$-channel exchanges are the $\Lambda$'s and $\Sigma$'s, $\alpha_{\mbox{\scriptsize{eff}}} \approx \alpha_{\Lambda} + \alpha_{\Sigma} \approx -1.4$ at $u \to 0$~\cite{collins}, which leads to a scaling power of $\approx -4.8$. Therefore, it is possible that the scaling power is steeper than $s^{-2}$. 

\subsection{Recoil polarization}

\begin{figure*}
\begin{center}
\subfigure[]{
{\includegraphics[angle=90,width=3.4in]{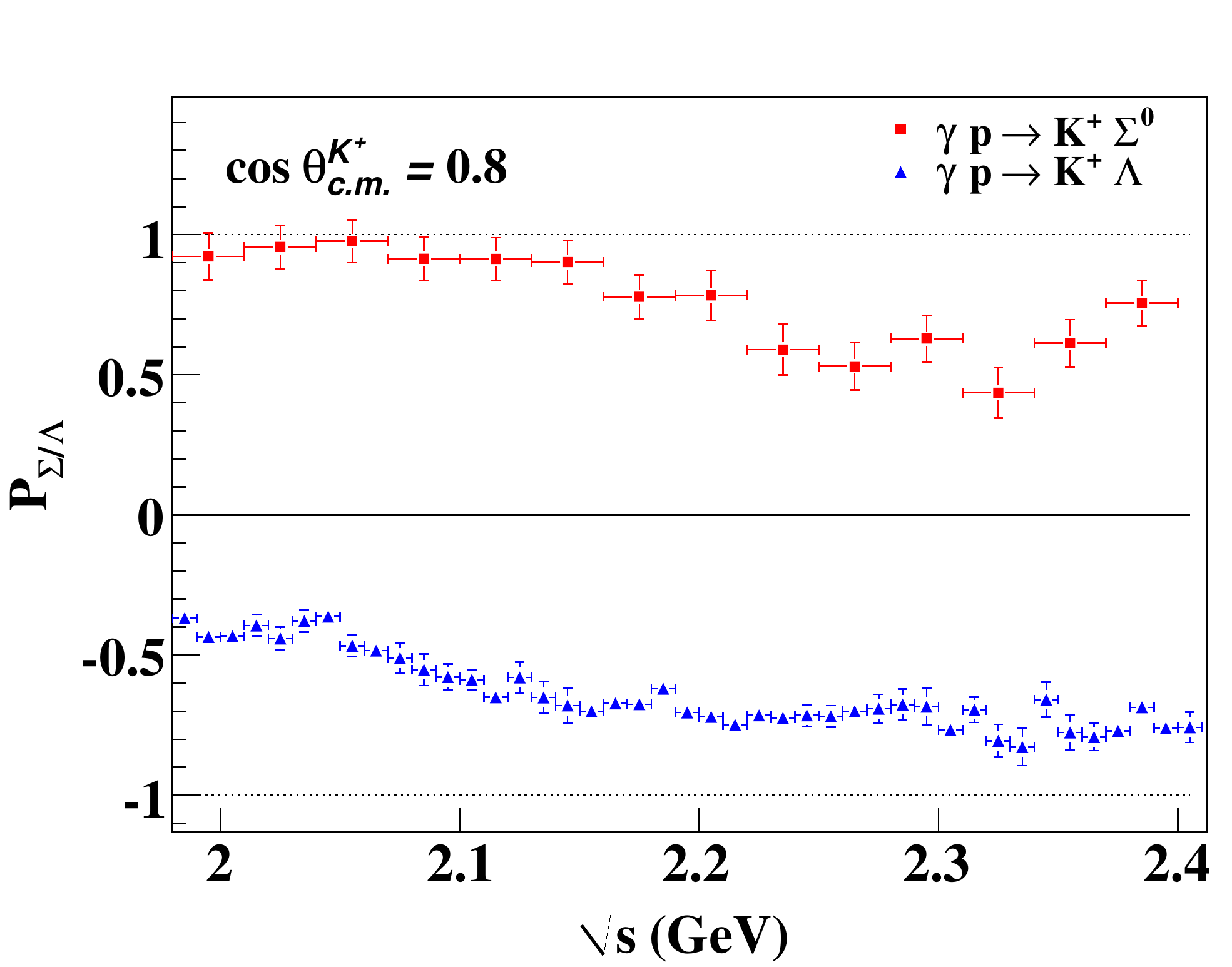}}
}
\subfigure[]{
{\includegraphics[angle=90,width=3.4in]{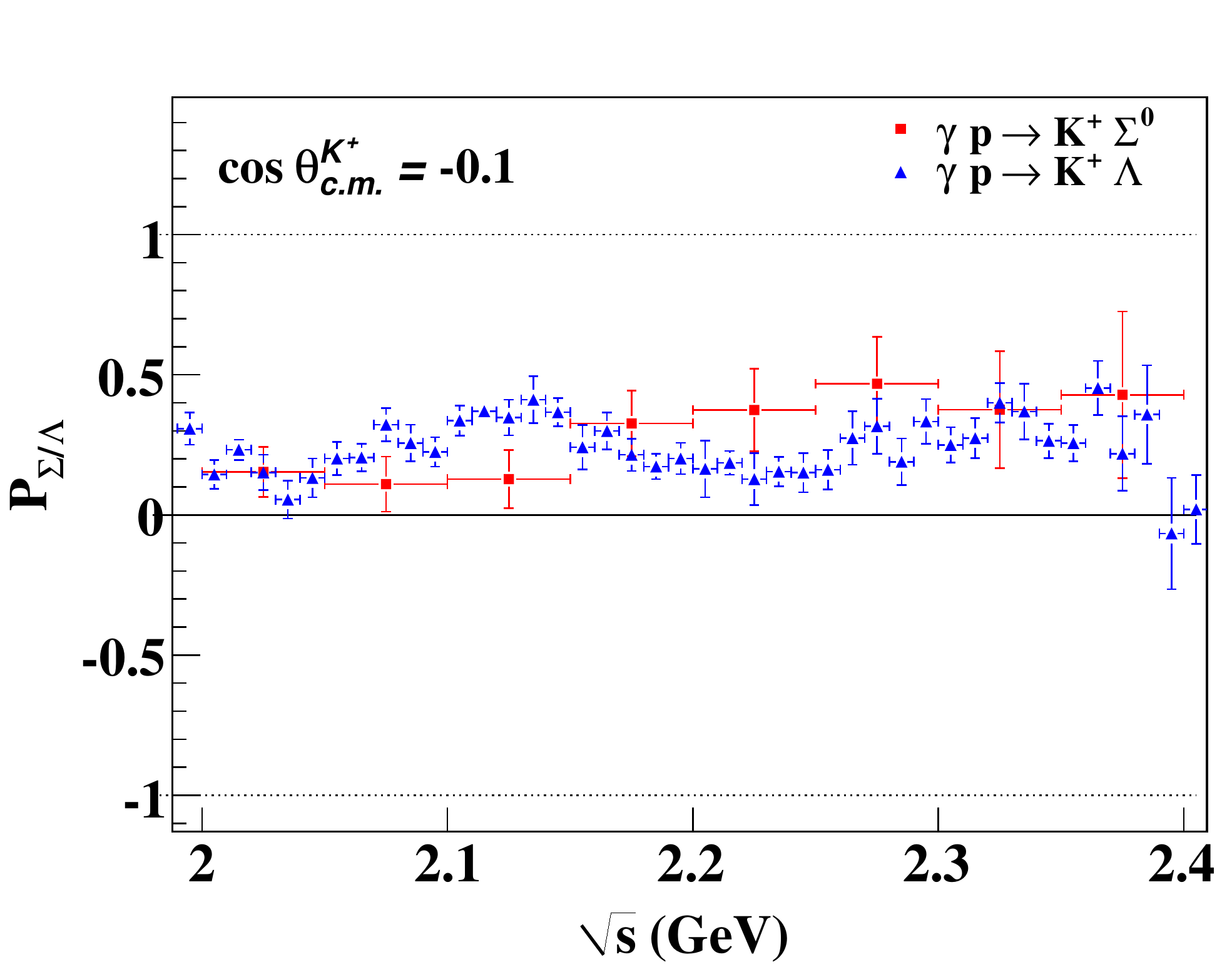}}
}
\caption[]{(Color online) 
The $SU(6)$ prediction of $P_\Sigma \approx -P_\Lambda$ is seen to hold at (a) forward-angles, but is broken for certain (b) mid- and backward-angle kinematics. The $P_\Lambda$ values are taken from Ref.~\cite{klam_prc}.}
\label{fig:klam_ksig_pol}
\end{center}
\end{figure*}

The overall trend of the polarization seems to be that $P_\Sigma$ is large and positive in the forward-angles and tends toward zero or negative values in the backward-angles. Many local structures are visible, especially in the backward-angle bins, possibly from resonance contributions, though the variations are smoother than seen in $K^+ \Lambda$~\cite{klam_prc}. In the static quark model, assuming an approximate $SU(6)$ symmetry~\cite{griffiths}, the spin-flavor configurations of the two hyperons are $|\Lambda^\uparrow\rangle = |u^\uparrow d^\downarrow s^\uparrow\rangle$ and $|\Sigma^{0\;\uparrow}\rangle = |u^\uparrow d^\uparrow s^\downarrow\rangle$. Therefore, it follows that $P_\Sigma \approx - P_\Lambda$. A new feature that we see from the present results is that this prediction is explicitly broken in certain kinematic regions. Fig.~\ref{fig:klam_ksig_pol} shows this for one region, where $P_\Lambda$~\cite{klam_prc} and $P_\Sigma$ are both non-zero and have the same sign. Similar features are visible in several other mid- and backward-angle bins, but not in any of the forward-angle bins. In other words, the $SU(6)$ prediction is not observed globally. Of course, $SU(6)$ is known to be a broken symmetry, and it is an interesting question by itself, as to why the $P_\Sigma \approx -P_\Lambda$ prediction seems to hold at high $\sqrt{s}$ and forward-angles. A possible answer may lie in the fact that the static quark model assumes that the production mechanisms for both hyperons are the same. This hypothesis no longer holds if $\Delta^\ast$ resonances contribute to the $K^+ \Sigma^0$ production, and $SU(6)$ can be broken explicitly.


\section{\label{section:conc} Summary}

We have presented high statistics measurements of differential cross sections and recoil polarizations for the reaction $\gamma p \rightarrow K^+ \Sigma^0$ from production threshold to $\sqrt{s}=2.84$~GeV with a wide angular coverage, using the CLAS detector at Jefferson Lab (electronic versions of the data can be found in Ref.~\cite{clasdb}). These new results significantly extend the previous $K^+ \Sigma^0$ world data on two separate fronts. Firstly, these precision polarization measurements will place additional constraints on future theoretical modeling of this reaction, and thereby help remove some of the ambiguities that presently plague this field. We find that the $SU(6)$ prediction of $P_\Sigma \approx -P_\Lambda$ is explicitly broken in some of the mid- and backward-angle bins, although it seems to be valid in the forward-angle bins. A possible explanation could be that for certain kinematic regions, the $K^+\Lambda$ and $K^+\Sigma^0$ productions proceed via different reaction mechanisms; however, this needs to be better understood.  
Secondly, the 300-MeV extension in energy coverage improves our understanding of the role non-resonant processes play in the reaction mechanism. Our data demonstrate that there is a significant $u$-channel contribution for this reaction, and is consistent with $u$-channel Regge exchanges of $\Lambda$ and $\Sigma$ hyperons. The forward-angle region is mostly dominated by $t$-channel processes. In the mid- and backward-angle regions, $s$-channel and $u$-channel amplitudes are expected to dominate, and these will interfere with each other. Therefore, it is important to cleanly separate the $u$-channel contribution before claiming the presence of any $s$-channel resonances. 

In a forthcoming work, we will present a coupled-channel partial-wave analysis incorporating the present work and the latest CLAS results for the $K^+ \Lambda$~\cite{klam_prc} and $\eta p$/$\eta' p$~\cite{eta_prc} channels. Each of these reactions has a relatively high strangeness content and is therefore expected to couple to a similar set of $s$-channel resonances. Additionally, all four analyses come from same data set and use similar analysis techniques, which should better keep systematic uncertainties under control as well. 

\begin{acknowledgments}
The authors thank the staff and administration of the Thomas Jefferson National Accelerator Facility 
who made this experiment possible. This work was supported in part by the U.S. Department of Energy 
(under grant No. DE-FG02-87ER40315); the National Science Foundation; the Italian Istituto Nazionale 
di Fisica Nucleare; the French Centre National de la Recherche Scientifique; the French Commissariat 
\`{a} l'Energie Atomique; the U.K. Research Council, S.T.F.C.; and the National Research Foundation of 
Korea. The Southeastern Universities Research Association (SURA) operated Jefferson Lab under United 
States DOE contract DE-AC05-84ER40150 during this work.
\end{acknowledgments}


\end{document}